\documentclass[fleqn,usenatbib]{mnras}

\usepackage{newtxtext,newtxmath}

\usepackage[T1]{fontenc}

\DeclareRobustCommand{\VAN}[3]{#2}
\let\VANthebibliography\thebibliography
\def\thebibliography{\DeclareRobustCommand{\VAN}[3]{##3}\VANthebibliography}

\usepackage{graphicx}	
\usepackage{amsmath} 
\usepackage{multirow}

\newcommand{\lr}{\ifmmode{\lambda_{R_e}}\else{$\lambda_{R_e}$}\fi}
\newcommand{\lre}{\ifmmode{\lambda_{R_{\rm{e}}}}\else{$\lambda_{R_{\rm{e}}}$}\fi}
\newcommand{\kms}{\ifmmode{\,\rm{km}\, \rm{s}^{-1}}\else{$\,$km$\,$s$^{-1}$}\fi}
\newcommand{\update}[1]{{#1}}
\newcommand{\updatetwo}[1]{#1}

\newcommand{\sextractor}{\textsc{sextractor}}
\newcommand{\galfit}{\textsc{galfit}}
\newcommand{\logm}{\ifmmode{\log_{10}(M_*/M_{\odot})}\else{$\log_{10}(M_*/M_{\odot})$}\fi}

\title[Galaxy Features and Kinematics]{The SAMI Galaxy Survey: Using  Tidal Streams and Shells to Trace the Dynamical Evolution of Massive Galaxies}

\author[T. H. Rutherford et al.]{
Tomas H. Rutherford,$^{1,2}$\thanks{E-mail: trut2989@uni.sydney.edu.au}
Jesse van de Sande,$^{1,2}$
Scott M. Croom,$^{1,2}$
Lucas M. Valenzuela,$^{3}$\newauthor
$\text{ }$Rhea-Silvia Remus,$^{3}$
Francesco D'Eugenio,$^{4,5}$
Sam P. Vaughan,$^{2,6,7,8}$
Henry R. M. Zovaro,$^{9,2}$\newauthor
$\text{ }$Sarah Casura,$^{10}$
Stefania Barsanti,$^{9,2,1}$
Joss Bland-Hawthorn,$^{1,2}$
Sarah Brough,$^{11,2}$
Julia J. Bryant,$^{1,2,12}$\newauthor
$\text{ }$Michael Goodwin,$^{13}$
Nuria Lorente,$^{13}$
Sree Oh,$^{14,9,2}$
Andrei Ristea$^{15,2}$
\\
$^{1}$Sydney Institute for Astronomy, School of Physics, A28, The University of Sydney, NSW, 2006, Australia\\
$^{2}$ARC Centre of Excellence for All Sky Astrophysics in 3 Dimensions (ASTRO 3D), Australia\\
$^{3}$Universit\"{a}ts-Sternwarte, Fakult\"{a}t f\"{u}r Physik, Ludwig-Maximilians-Universit\"{a}t M\"{u}nchen, Scheinerstr. 1, 81679 M\"{u}nchen, Germany\\
$^{4}$Kavli Institute for Cosmology, University of Cambridge, Madingley Road, Cambridge, CB3 0HA, UK\\
$^{5}$Cavendish Laboratory - Astrophysics Group, University of Cambridge, 19 JJ Thomson Avenue, Cambridge, CB3 0HE, UK\\
$^{6}$School of Mathematical and Physical Sciences, Macquarie University, NSW 2109, Australia\\
$^{7}$Astronomy, Astrophysics and Astrophotonics Research Centre, Macquarie University, Sydney, NSW 2109, Australia\\
$^{8}$Centre for Astrophysics and Supercomputing, School of Science, Swinburne University of Technology, Hawthorn, VIC 3122, Australia\\
$^{9}$Research School of Astronomy and Astrophysics, Australian National University, Canberra, ACT 2611, Australia\\
$^{10}$Hamburger Sternwarte, Universit\"{a}t Hamburg, Gojenbergsweg 112, 21029 Hamburg, Germany\\
$^{11}$School of Physics, University of New South Wales, NSW 2052, Australia\\
$^{12}$Astralis-USydney, School of Physics, University of Sydney, NSW 2006, Australia\\
$^{13}$AAO-MQ, Faculty of Science \& Engineering, Macquarie University. 105 Delhi Rd, North Ryde, NSW 2113, Australia\\
$^{14}$Department of Astronomy and Yonsei University Observatory, Yonsei University, Seoul 03722, Republic of Korea\\
$^{15}$International Centre for Radio Astronomy Research, The University of Western Australia, 35 Stirling Highway, Crawley WA 6009, Australia\\
}

\date{Accepted 2024 January 30. Received 2024 January 23; in original form 2023 August 17}

\pubyear{2023}

\begin{document}
\label{firstpage}
\pagerange{\pageref{firstpage}--\pageref{lastpage}}
\maketitle

\begin{abstract}
Slow rotator galaxies are distinct amongst galaxy populations, with simulations suggesting that a mix of minor and major mergers are responsible for their formation. A promising path to resolve outstanding questions on the type of merger responsible, is by investigating deep imaging of massive galaxies for signs of potential merger remnants. We utilise deep imaging from the Subaru-Hyper Suprime Cam Wide data to search for tidal features in massive ($\logm > 10$) early-type galaxies (ETGs) in the SAMI Galaxy Survey. We perform a visual check for tidal features on images where the galaxy has been subtracted using a Multi-Gauss Expansion (MGE) model. We find that $31${\raisebox{0.5ex}{\small$^{+2}_{-2}$}} percent of our sample show tidal features. When comparing galaxies with and without features, we find that the distributions in \update{stellar mass,} light-weighted mean stellar population age and H$\upalpha$ equivalent width are significantly different, whereas spin (\lre), ellipticity and bulge to total ratio have similar distributions. When splitting our sample in age, we find that galaxies below the median age (10.8 Gyr) show a correlation between the presence of shells and lower \lre, as expected from simulations. We also find these younger galaxies which are classified as having "strong" shells have lower \lre. However, simulations suggest that merger features become undetectable within $\sim 2-4$ Gyr post-merger. This implies that the relationship between tidal features and merger history disappears for galaxies with older stellar ages, i.e. those that are more likely to have merged long ago.
\end{abstract}




\section{Introduction}

Galaxy mergers play an important role in the hierarchical structure formation theory of the Universe \citep{1978MNRAS.183..341W}, being a key aspect of mass build-up in the $\Lambda$CDM paradigm. While the overall picture is well established, the impact of mergers on specific sub-classes or individual galaxies is less well understood. Slow rotator galaxies are a subset of galaxies characterised by a low spin parameter (\lre) \citep{2007MNRAS.379..401E}, large stellar mass and old stellar ages. However, it is not clear which processes cause the morphological (spin-down) and quenching transformation of these galaxies \citep[e.g.][]{2022MNRAS.509.4372L}. Evidence from simulations \citep[e.g.,][]{2009A&A...501L...9D, 2009MNRAS.397.1202J, 2011MNRAS.416.1654B, 2014MNRAS.444.3357N, 2017ApJ...837...68C, 2017MNRAS.464.3850L, 2018MNRAS.473.4956L, 2017MNRAS.468.3883P, 2020MNRAS.493.3778S,2020IAUFM..30A.208L} suggests that galaxy mergers are capable of kinematically transforming these galaxies. However when these mergers happen, the ratio of minor to major mergers and the importance of gas are still outstanding questions. Indeed, some observational studies have found no link between merger signatures and slow rotators \citep[e.g.,][]{2016ApJ...832...69O}.

Galaxy interactions and mergers leave remnants of their existence in the form of tidal features, long understood to be relics of past encounters \citep[e.g.,][]{1972ApJ...178..623T,2005AJ....130.2647V,2018ApJ...857..144H,2019A&A...632A.122M,2022ApJS..262...39H}. Tidal tails and streams are formed when material is stripped from a primary gas-rich, disc-dominated galaxy or from a low-mass companion during interactions \citep{1992AJ....103.1089B,2008ApJ...683...94O, 2018ApJ...857..144H}. Simulations have suggested that streams indicate a circular infall of a companion (large impact parameter), and shells a nearly complete radial infall \citep{2019MNRAS.487..318K}. However, it should be noted that the evolution from a radial merger to a shell is not one-to-one, as radial infall mergers can also form other structures such as diffuse fans \citep{1997ApJ...490..664W}. 

If slow rotators (SRs) are formed through galaxy mergers \citep{2022MNRAS.509.4372L}, we expect to see a higher fraction of recent merger features around these slow rotator galaxies. The galaxy spin parameter \lre\ has traditionally been used to classify galaxies as slow rotators ($\lre \lesssim 0.1$) or fast rotators ($\lre >0.1$) \citep{2007MNRAS.379..401E}. \lre\ is defined as $\lre = {\langle R|V|\rangle} / {\langle R \sqrt{V^2 + \sigma^2}\rangle}$, a luminosity-weighted rotational velocity, normalised by the second velocity moment $V^2+\sigma^2$. \lre\ can be interpreted as a measure of how rotationally or dispersion supported a galaxy is. More recently, slow rotators have been defined within a region of \lre-$\varepsilon$ (spin-ellipticity) space \citep[e.g.,][]{2011MNRAS.414..888E, 2016ARA&A..54..597C, 2021MNRAS.505.3078V}, with low \lre\ ($\lesssim 0.2$) and low $\varepsilon$ ($\leq0.4$).

The role of major versus minor mergers in the evolution of slow rotators is not yet clear from simulations. Whilst major mergers have long been believed to be essential for spinning-down galaxies, it is also known that processes such as  minor mergers, secular evolution, fly-by encounters, harassment and dynamical friction can impact and even dominate spin evolution \citep{2017ApJ...837...68C}. Indeed, slow rotators can form from a series of minor mergers \citep{2020MNRAS.493.3778S}, and this may even be the dominant driver of morphological transformations over cosmic time \citep{2017ApJ...837...68C, 2017MNRAS.465.2895L}. \lre\ has been shown to decrease in simulations and observations from a maximum at $z=1$ \citep{2018ApJ...858...60B}, however, this is independent of merger history in the case of simulations \citep{2020MNRAS.494.5652W}. \citet{2017ApJ...837...68C} and \citet{2017MNRAS.465.2895L} argue that major mergers are not the main driver of the spin-down of galaxies over cosmic time. 

In order to confidently link tidal features to a potential merger history, an understanding of the timescale on which these features fade to be undetectable, or settle back into the main body of the galaxy is required. Post-merger galaxies can mostly only be identified via their tidal features during this timescale, which may differ from the timescale for \lre to settle post-merger. Generally, tidal features are visible for a few Gyr \citep[e.g.,][]{2008MNRAS.391.1137L, 2010MNRAS.404..575L,2010MNRAS.404..590L,2017MNRAS.465.2895L,2019A&A...632A.122M,2021ApJ...912...45N}, although various studies have a large variation in timescales. The best combination of techniques can make detection possible up to $\lesssim$ 2 Gyr \citep{2021ApJ...912...45N}, or potentially $\lesssim 4$ Gyr \citep{2018MNRAS.480.1715P}. However, these times have been shown to be dependent on gas fraction \citep{2018ApJ...857..144H}, decreasing to $\leq 300$ Myr for low gas fractions ($\sim 20 \%$), and reaching up to $\geq 1$ Gyr for higher gas fractions ($\sim 50\%$) \citep{2010MNRAS.404..575L}. \citet{2022ApJS..262...39H} estimate that for massive early-type galaxies (ETGs), the lifetime of tidal features is $\sim 3$ Gyr. It should be noted, however, that these timescales are often related to identifying a merger in progress \citep[e.g.,][]{2010MNRAS.404..590L, 2010MNRAS.404..575L, 2019ApJ...872...76N, 2021ApJ...912...45N}, and the timescale for tidal features at large radii to exist may be significantly longer. Additionally, different classes of features may have different detectability timescales.  \citet{2019A&A...632A.122M} investigated the lifetime of different types of features. They estimated a survival time of $\sim 2$ Gyr for tidal tails, $\sim 3$ Gyr for streams, and $\sim 4$ Gyr for shells.

The detection of tidal features is also strongly dependent on surface brightness limits. \cite{2014A&A...566A..97J} found that the detection time was on average $\sim 2$ times as long for a surface brightness limit of 28 mag arcsec$^{-2}$ as compared to 25 mag arcsec$^{-2}$. They found environment had an important effect, where a cluster potential was able to strip merger features and reduce the detection time. \cite{2022MNRAS.513.1459M} also found that detection rates depend on surface brightness limits, with 80 per cent of flux in features around Milky Way like galaxies identified at 30-31 mag arcsec$^{-2}$, falling to 60 per cent at 29.5 mag arcsec$^{-2}$.

The vast array of different merger scenarios can have a strong impact on the evolution of the galaxy, with the importance of gas specifically impacting both the kinematic evolution and star formation quenching. Whilst mergers may kinematically spin-down a galaxy \citep[e.g.,][]{2014MNRAS.444.3357N, 2020IAUFM..30A.208L} and leave an SR remnant, a fast-rotating disc can be subsequently formed post merger. However, the formation of a fast-rotating disc post-merger is dependent on the gas fraction \citep{2020IAUFM..30A.208L}, where wet mergers can even spin-up a galaxy post-merger, via newly-formed stars of high rotational speed. Mergers can also cause quenching in galaxies provided there is strong active galactic nuclei (AGN) feedback \citep{2017MNRAS.470.3946S}, but this strong feedback is generally only induced by the gaseous disc being disrupted \citep{2017MNRAS.465..547P}. Dry mergers tend to spin-down and morphologically disrupt galaxies, and wet mergers are capable of inducing star formation in a new disk. If this disk has enough stellar mass to counteract the stars sent to hotter orbits when the merger occurred, there can be little to no change in \lre.

The identification and analysis of galaxy mergers from observations is complicated by the low surface brightness of the tidal features, and the large radii at which they often occur. Studies using simulations often analyse \lre\ and studies using observations perform visual merger checks, and thus work that bridges this gap is a key part of identifying the impact of galaxy mergers on slow rotator formation. Recent work by \citet{2022arXiv220808443V}, building on tidal feature classification by \citet{2020MNRAS.498.2138B} on the MATLAS sample \citep{2015MNRAS.446..120D, 2020MNRAS.491.1901H} has shown a correlation between shells and low \lre\ values. 

In this paper, we use the recent wealth of deep optical imaging from the Subaru Hyper-Suprime Cam (HSC) to expand on the MATLAS (Mass Assembly of early-Type GaLAxies with their fine Structures) work with a significantly larger sample size. We use visual identification of tidal features in model-subtracted massive galaxies (e.g., shells, tidal streams, tails) from HSC-SSP \citep{2019PASJ...71..114A} in addition to standard cutout feature visual identification, complemented by kinematic data from the SAMI Galaxy Survey \citep{2021MNRAS.505..991C}. We aim to address some of the questions surrounding the role of galaxy mergers in the formation pathway of slow rotators.

Our work is structured as follows. Section 2 discusses the data used in this paper. Section 3 describes our method. Section 4 states the results. Section 5 discusses these results in context with previous work. Section 6 presents a conclusion to this work. Throughout this paper, we adopt a $\Lambda$CDM cosmology, with $H_0=70$ km s$^{-1}$ Mpc$^{-1}$, $\Omega_m=0.3$, $\Omega_{\Lambda} = 0.7$. We further assume a Chabrier \citep{2003PASP..115..763C} stellar initial mass function (IMF), and an AB magnitude system \citep{1983ApJ...266..713O}.

\section{Data}
\label{sec:data}
\subsection{The SAMI Survey}
The SAMI instrument \citep{2012MNRAS.421..872C} provides a 1 degree diameter field of view and is mounted on the Anglo-Australian Telescope. SAMI uses 13 fibre bundles \citep[Hexabundles;][]{2011OExpr..19.2649B,2014MNRAS.438..869B}, each having a 75\% fill factor. There are 61 fibres of 1.6\arcsec diameter within each bundle, giving a diameter of 15\arcsec to each IFU. The IFUs and 26 sky fibres are fed to the AAOmega spectrograph \citep{2006SPIE.6269E..0GS}. The 580V grating at 3570-5750\AA\ is used, giving a resolution of R=1808 ($\sigma$=70.4 \kms), as well as the 1000R grating from 6300-7400\AA, which gives a resolution of R=4304 ($\sigma$=29.6 \kms) \citep{2017ApJ...835..104V}. 

The SAMI Galaxy survey \citep{2012MNRAS.421..872C,2015MNRAS.447.2857B} target selection came from the GAMA \citep{2011MNRAS.413..971D} survey, as well as eight low-redshift clusters \citep{2017MNRAS.468.1824O}. Reduced data cubes \citep{2015MNRAS.446.1551S} are available with the SAMI Galaxy Survey data releases \citep{2015MNRAS.446.1567A,2018MNRAS.475..716G,2018MNRAS.481.2299S,2021MNRAS.505..991C}, as well as stellar kinematic maps.

Our sample contains \lre\ values derived from spatially resolved kinematic measurements, detailed in \citet{2017ApJ...835..104V}, with both an aperture correction \citep{2017MNRAS.472.1272V}, and a seeing correction \citep{2020MNRAS.497.2018H, 2021MNRAS.505.3078V}. $R_e$ is defined as the semimajor axis effective radius of a galaxy, and the ellipticity $\varepsilon$ is defined in terms of the axis ratio, $b/a = 1-\varepsilon$. $R_e$ and $\varepsilon$ values are taken from MGE fits derived by \cite{2021MNRAS.504.5098D}, using code from \cite{2002MNRAS.333..400C}. Stellar mass measurements are taken from \cite{2015MNRAS.447.2857B}, using Milky-Way-extinction-corrected apparent $g$ and $i$ magnitudes, using the technique from \cite{2011MNRAS.418.1587T}.

Luminosity-weighted age measurements were derived by \citet{2022MNRAS.516.2971V}. The \textsc{ppxf} code \citep{2004PASP..116..138C, 2017MNRAS.466..798C} was used to fit MILES single-age, single-metallicity stellar populations (SSP) models \citep{2015MNRAS.449.1177V}. The SSP models used templates from \citet{2004ApJ...612..168P,2006ApJ...642..797P}. Bulge to total flux ratios were derived by \citet{2022MNRAS.516..942C}, using the \textsc{profound} \citep{2018MNRAS.476.3137R} and \textsc{profit} \citep{2017MNRAS.466.1513R} codes, applied to r-band KiDS (Kilo-Degree Survey) DR4.0 photometry \citep{2019A&A...625A...2K}. In our analysis, we begin by including bulge to total ratios of galaxies that have an appropriate two component fit\footnote{To have an appropriate fit, a galaxy was required to have the flag R\_N\_COMP = 2, i.e. that the recommended number of S\'{e}rsic components was 2. For more detail, see \cite{2022MNRAS.516..942C}.}. We then extend this, by assigning bulge to total ratios of 1 to one component fit galaxies with S\'{e}rsic indices above 2.5, and bulge to total ratios of 0 to one component fit galaxies with S\'{e}rsic indices below 1. This allows us to include pure ellipticals and disks which cannot be fit with two components in our analysis. 

For each galaxy, the H$\upalpha$ equivalent width (H$_{\text{H}\upalpha}$) was measured within a circular apertures with radius $1R_e$, where we adopted the MGE measurements from SAMI DR3 \citep{2021MNRAS.504.5098D,2021MNRAS.505..991C}. The equivalent width is measured as the total H$\upalpha$ flux (not corrected for extinction) divided by the mean continuum level within the rest-frame wavelength range from 6500\,\AA\ to 65400\,\AA.

\subsection{The HSC SSP Survey}
Located on the summit of Maunakea in Hawaii, the wide-field imaging camera Hyper-Suprime-Cam (HSC) is mounted on the prime focus of the 8.2m Subaru telescope \citep{2018PASJ...70S...4A, 2015ApJ...807...22M}. HSC utilises the full 1.5 degree diameter field of view of the Subaru telescope, with 116 Hamamatsu Deep Depletion CCDs, with $2000 \times 4000$ pixels each. Each of the $15\mu m$ pixels covers $0.168\arcsec$ on the sky. There are five broad-band (\textit{grizy}) and 4 narrow band filters.

This paper uses the HSC $r$-band Wide data from HSC data release 2, which has a surface brightness limit of $27.8 \pm 0.5$ mag arcsec$^{-2}$, derived in Section \ref{sec:surf_bright}. We use the Wide data due to its overlap with the GAMA/SAMI sample, and data release 2 was shown by \cite{2022ApJS..262...39H} to be most suitable for extended, LSB objects. Image cutouts of $361\times 361$ pixels (approximately $60\arcsec \times 60\arcsec$), centred on the chosen galaxy, were used for model subtraction images. Greyscale inspection images were taken to be at least $10R_e$ in size, requiring cutouts of $829\times 829$ pixels (approximately $140\arcsec \times 140\arcsec$) as standard, with larger taken when necessary.

\subsection{Sample Selection}
We take our sample to be non-cluster SAMI ETGs from SAMI DR3 \citep{2021MNRAS.505..991C}, as these galaxies have HSC imaging coverage. Morphology measurements were taken from \citet{2016MNRAS.463..170C}, which used the classification method from \citet{2014MNRAS.439.1245K} on SDSS DR9 colour images \citep{2012ApJS..203...21A}. We selected all galaxies above $\logm = 10$ to balance our aim to investigate massive slow rotators, and to appropriately compare to \citet{2022arXiv220808443V}. Some galaxies in this sample are discarded due to issues with their imaging (artefacts impacting the galaxy image), or identifiable spiral arms in HSC-SSP despite their previous morphological classification from SDSS imaging. Not all galaxies have values for all parameters. This is due to several factors, most importantly low signal to noise and insufficient radial coverage. The final sample contains $411$ ETGs with good imaging, of which:
\begin{itemize}
    \item 325 galaxies have \lre\ values.
    \item 268 have B/T ratios.
    \item 405 have ellipticity values.
    \item 386 have light-weighted age measurements.
\end{itemize}

\begin{figure*}
    \centering
    \includegraphics[width=\textwidth]{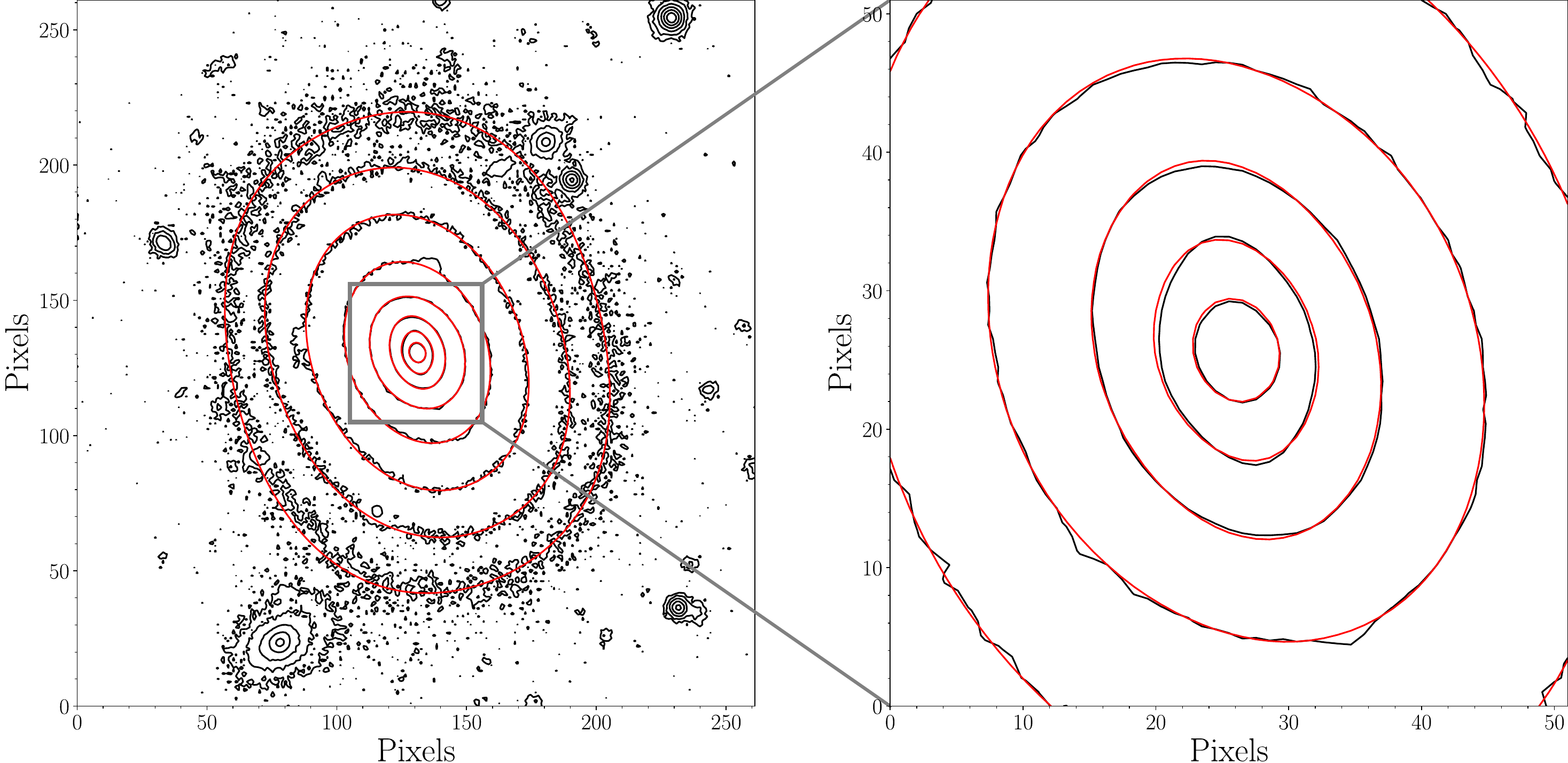}
    \caption{An example of an MGE model of galaxy with CATID 91697. The left panel has flux contour lines in black, with MGE model contour lines in red. Contour lines are spaced by 1 mag arcsec$^{-2}$. The grey square contains the central region of the galaxy, and is enlarged and shown in the right-hand panel. The pixel scale of all cutouts is 0.168 arcsec pixel$^{-1}$.}
    \label{fig:mge_fit}
\end{figure*}

\subsection{HSC Surface Brightness Limits}
\label{sec:surf_bright}
The identification of low surface brightness (LSB) features in galaxy surveys is limited by the minimum surface brightness of a survey. Previously unobserved features often appear in a deeper survey, and these features can be required to be several magnitudes brighter than the reported depth of a survey to be identified at a high completeness level \citep{2018ApJ...866..103K}.

However, it should be noted that the effect of surface brightness limits differs for different tidal features. \citet{2019A&A...632A.122M} found that the detection of streams is highly sensitive to surface brightness limits ($2-3\times$ more streams with a brightness cut of 33 mag arcsec$^{-2}$ compared to 29 mag arcsec$^{-2}$). There was no sensitivity dependence found for tails, and only a mild dependence for shells.

The various surveys on LSB features use different methods and aperture sizes to estimate the limiting surface brightness. \citet{2013ApJ...765...28A} note that differences as large as 2-3 mag arcsec$^{-2}$ can exist between surveys, making comparison extremely difficult. For completeness, we will describe here our derivation of our surface brightness limits.

Similarly to \citet{2013ApJ...765...28A}, we placed 40 circular apertures of area 1 arcsec$^2$ on empty fields of 20 HSC cutouts. The root-mean-square variations between the total sum of flux within each aperture, for each cutout, was calculated. These values were then converted to mag arcsec$^{-2}$ with Equation \ref{eq:flux_to_mag}, where FLUXMAG0 = 63095734448.0194.
\begin{align}
    \label{eq:flux_to_mag}
    SB_{\text{mag arcsec$^{-2}$}} = 2.5\times\log_{10}\bigg(\frac{\text{FLUXMAG0}}{\text{flux}}\bigg)
\end{align}
This resulted in a distribution of surface brightness limits, where we took the mean and standard deviation to be our limiting surface brightness, and its uncertainty. This estimation of our surface brightness limit was found to be $27.8 \pm 0.5$ mag arcsec$^{-2}$.
\begin{figure*}
    \includegraphics[width = \textwidth]{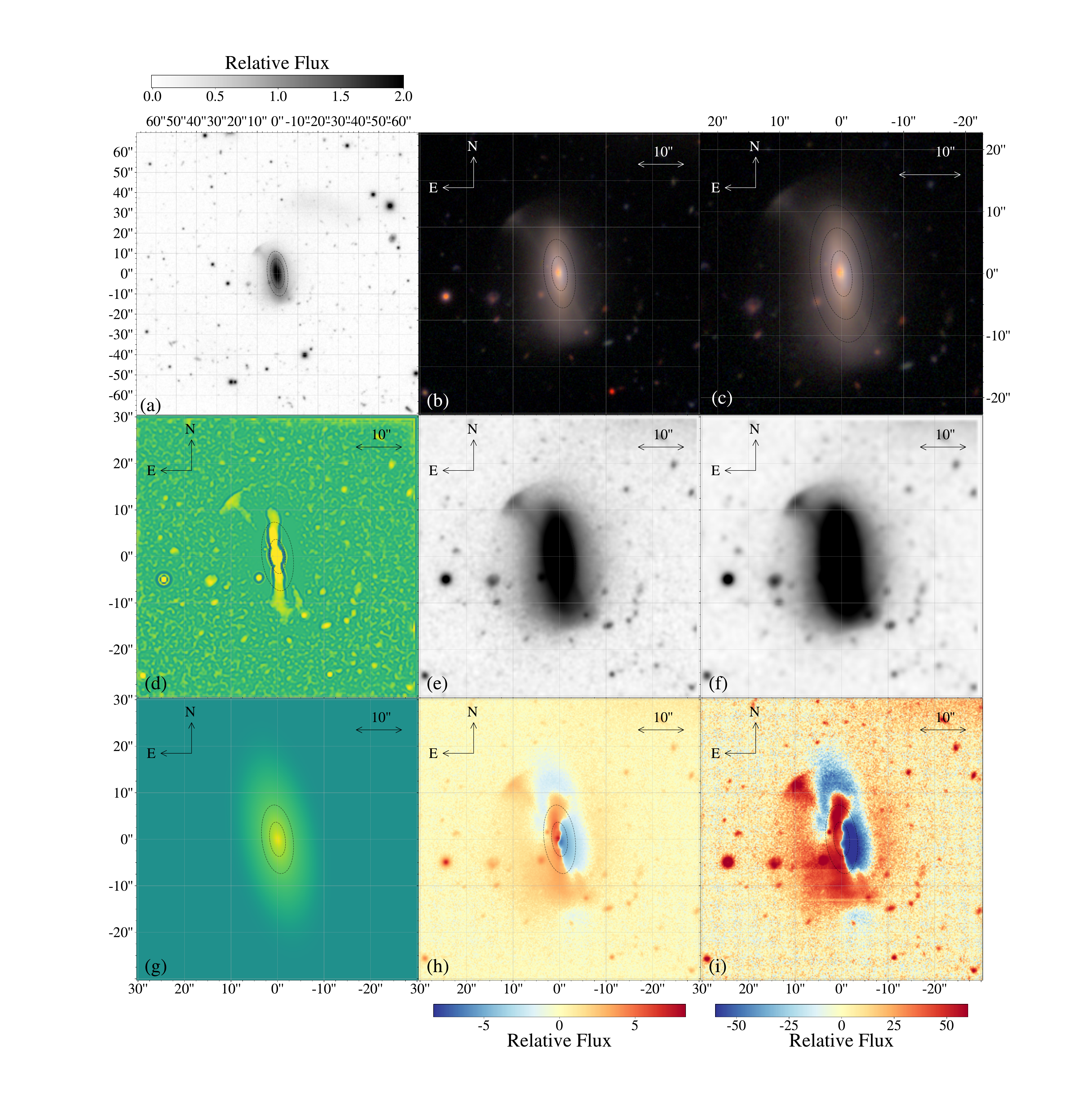}
    \caption{An example image (DEC vs RA) for analysis of galaxy with CATID 49734. Panel (a) contains the r-band cutout for the galaxy. Panels (b) and (c) contain an rgb image of the galaxy, and an enlarged version of the rgb image respectively. Panel (d) contains the r-band cutout for the galaxy, with the spatial frequency filters from \citet{2018ApJ...866..103K} applied,  and panels (e) and (f) contain r-band cutouts with increased contrast. Panel (g) contains an MGE model of the galaxy, panel (h) contains the MGE model-subtracted residual, and panel (i) contains the same residual divided by noise. The dashed ellipses represent the radii $1R_e$ and $2R_e$ (and $3R_e$ in panel (c)). A shell-like tidal feature can be seen to the upper left of the images.}
    \label{fig:residual_example_image}
\end{figure*}

\section{Method}
\label{sec:method}
\subsection{MGE Galaxy Profile Modelling}
The visual inspection and analysis of LSB tidal features in deep imaging is complicated by separating these features from the bright, smoothly varying flux of the main body of the galaxy. To overcome this issue, analysis of model-subtracted residual images has been employed in previous work \citep[e.g.,][]{2006ApJ...640..241B, 2008MNRAS.388.1537M, 2019MNRAS.486.2643M}. The modelling code \galfit\ \citep{2002AJ....124..266P} has traditionally been used to fit physically motivated Sérsic profiles for the required flux models. MGE is an alternate method, being significantly faster than \galfit\, with high-efficiency algorithms delivering a fit in $\sim 2$ minutes. The efficiency of MGE comes from the fitting of a number of sectors of a given flux profile, rather than pixel-by-pixel. Given our desire for fast model generation which can give us the smallest residuals after model subtraction, physical motivation is less important than efficiency, and we utilise MGE in this work.

We follow the MGE fitting code from \citet{2021MNRAS.504.5098D}. The code starts by using \sextractor\ \citep{1996A&AS..117..393B} to identify all sources in the HSC cutout. \sextractor\ provides an image mask to all sources other than the central galaxy. The  point-spread function (PSF) is characterised by fitting 2-5 stacked circular Gaussians.

Provided with a PSF and image mask, the galaxy flux is fit. We do not use regularised fits \citep{2009MNRAS.398.1835S}, which require the flattest Gaussian to have the roundest axial ratio which still reproduces observations \citep{2002MNRAS.333..400C}. Unregularised fits yield the lowest $\chi^2$ values and more realistic galaxy shapes \citep{2021MNRAS.504.5098D}. We present an example MGE fit in Figure \ref{fig:mge_fit}.

The MGE modelling code used in this work \citep{2002MNRAS.333..400C} fits an integer pixel to the galaxy centre. As a result, the flux in the centre of a galaxy is usually not fit as well as the outer regions, and thus model subtraction near the galaxy centre is poor. This can be seen in examining panel (i) in Figure \ref{fig:residual_example_image}. However, as we expect tidal features at large radii, this sub-pixel shift does not affect the identification of these features.

\subsection{Model Subtraction and Visual Inspection}

The visual inspection of each galaxy for tidal features utilised an image with multiple panels. Each image includes: $r$-band image, rgb image, MGE model, model-subtracted residuals, and relative model-subtracted residuals. An example image can be seen in Figure \ref{fig:residual_example_image}. Panel (a) shows an r-band cutout for the galaxy. Panels (b) and (c) show an rgb image for the galaxy. Panel (d) shows the r-band cutout, with spatial frequency filters from \citet{2018ApJ...866..103K}, with the goal of making non-spherically-symmetric features more apparent. Panels (e) and (f) show increased contrast versions of the r-band cutout. Panel (g) shows the MGE model for the galaxy, with panels (h) and (i) showing the model subtracted residual, and model subtracted residual divided by uncertainty respectively.

Tidal features are often divided into classes, based on the physical process they formed from \citep[e.g.,][]{2013ApJ...765...28A,2020MNRAS.498.2138B,2023arXiv230704967D}. We divide these into two for this work: tidal streams, and shells. Tidal streams are seen as "streams" of stars, appearing to move radially inwards to the galaxy, and in some cases the satellite galaxies associated with these streams are still visible. We acknowledge here that there is significant ambiguity between a shell, stream, ring and a weak spiral arm in LSB imaging. Although our sample was selected to contain only ETGs, the morphological classifications used were taken from visual inspection of SDSS imaging. The increased depth of HSC imaging allows for visual identification of rings and spiral arms previously unseen. We defined a shell as a feature that curves around a galaxy with approximately constant radius, and unlike most spiral arms or rings, is separate from the main body of the galaxy. We made a point to only classify galaxies without knowing their \lre\ values, to avoid any unconscious bias. This also meant we did not downselect our galaxies after re-examination. There are some fast rotating disc galaxies with weak shell features that could plausibly be a ring within the disc, rather than a shell. We accept this as a reality of our method, which cannot be easily reduced without biasing our sample (examples can be seen in Figure \ref{fig:interesting_gals}).

Each HSC model-subtracted residual image was inspected individually by a group of three people (THR, JvdS, SMC). Each person was able to individually decide if an image had any stream or shell-like tidal features, and give a 1-5 value for the "strength" of that feature. This strength classification represented the confidence a classifier had in a feature existing, and often correlated with how bright and separated from the main galaxy body the feature was. When inspecting the image seen in Figure \ref{fig:residual_example_image}, one could focus on the top left panel, panel (a). The dynamic range of the greyscale image could be changed interactively for each galaxy, allowing for the user to select a unique scaling to best identify features for a given galaxy. A feature was determined to be significant if at least two people identified it. The strength was taken to be the average strength selected.

Once individual classifications were complete, a set of galaxies which had disagreements in strength of $\geq 2$ were selected, and inspected as a group. Disagreements were identified and a group classification was agreed upon. Most disagreements were due to ambiguity between a shell and a stream. We note that other galaxy observables, such as metallicity, kinematics and HI maps, can also be used alongside visual features to identify past interactions \citep[e.g.,][]{2010A&A...521A..63L}. However, this work is focused on correlating these observables to tidal features and thus we will not be using this technique.

\section{Results}
\label{sec:results}
\subsection{Merger/Tidal Features in the SAMI Galaxy Survey}
\begin{figure*}
	\includegraphics[width=\textwidth]{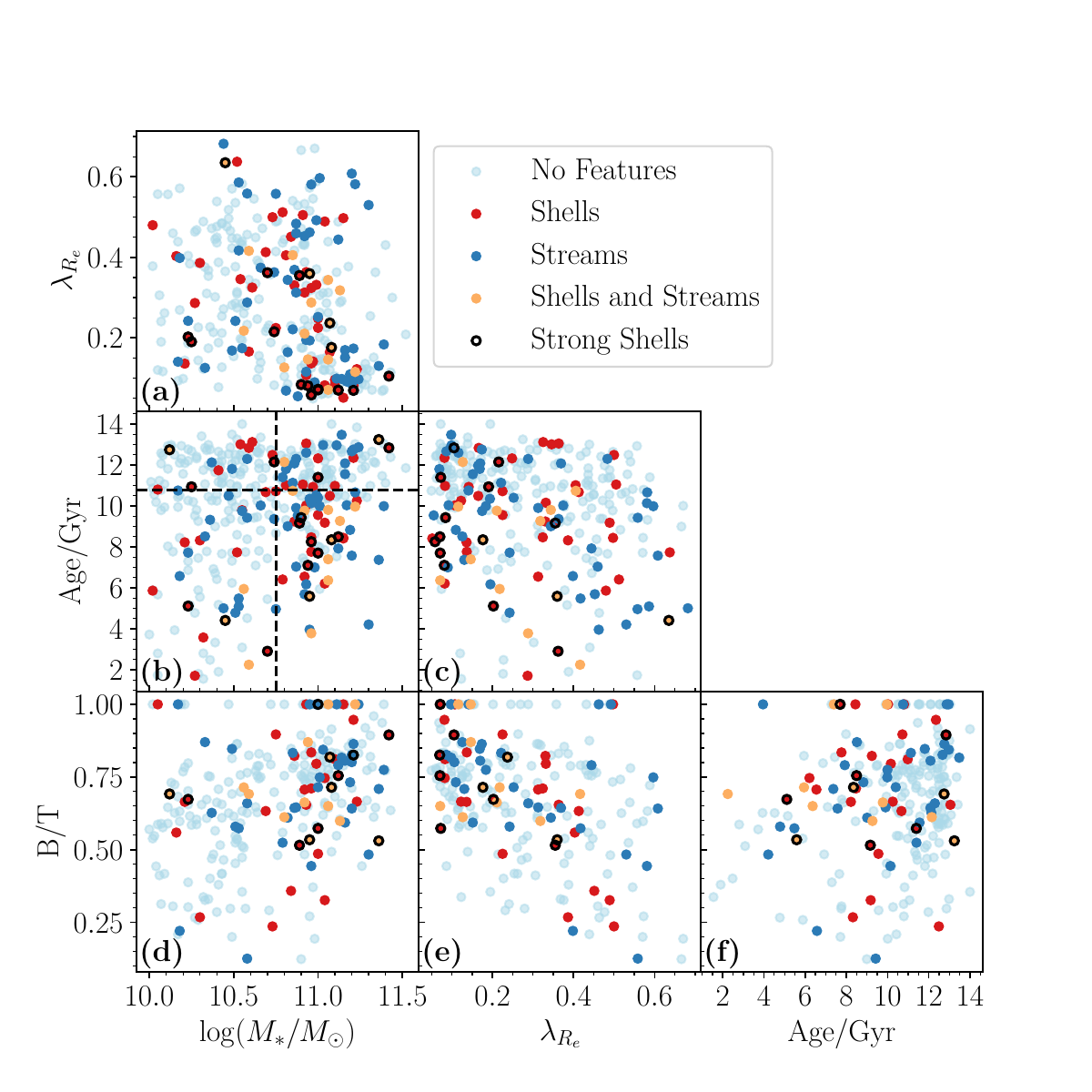}
    \caption{Distributions of our sample galaxies in all relevant parameter spaces, using the parameters \lre, \logm, Age and B/T. Galaxies with no features are coloured in sky blue, galaxies with shells are coloured in red, galaxies with streams are coloured in dark blue, and galaxies with streams and shells are coloured in black. Shell galaxies classified with a strength of at least 3 are circled in black. The median stellar mass and age are shown as dashed lines in panel (b).}
    \label{fig:all_parms}
\end{figure*}
\begin{figure*}
	\includegraphics[width=\textwidth]{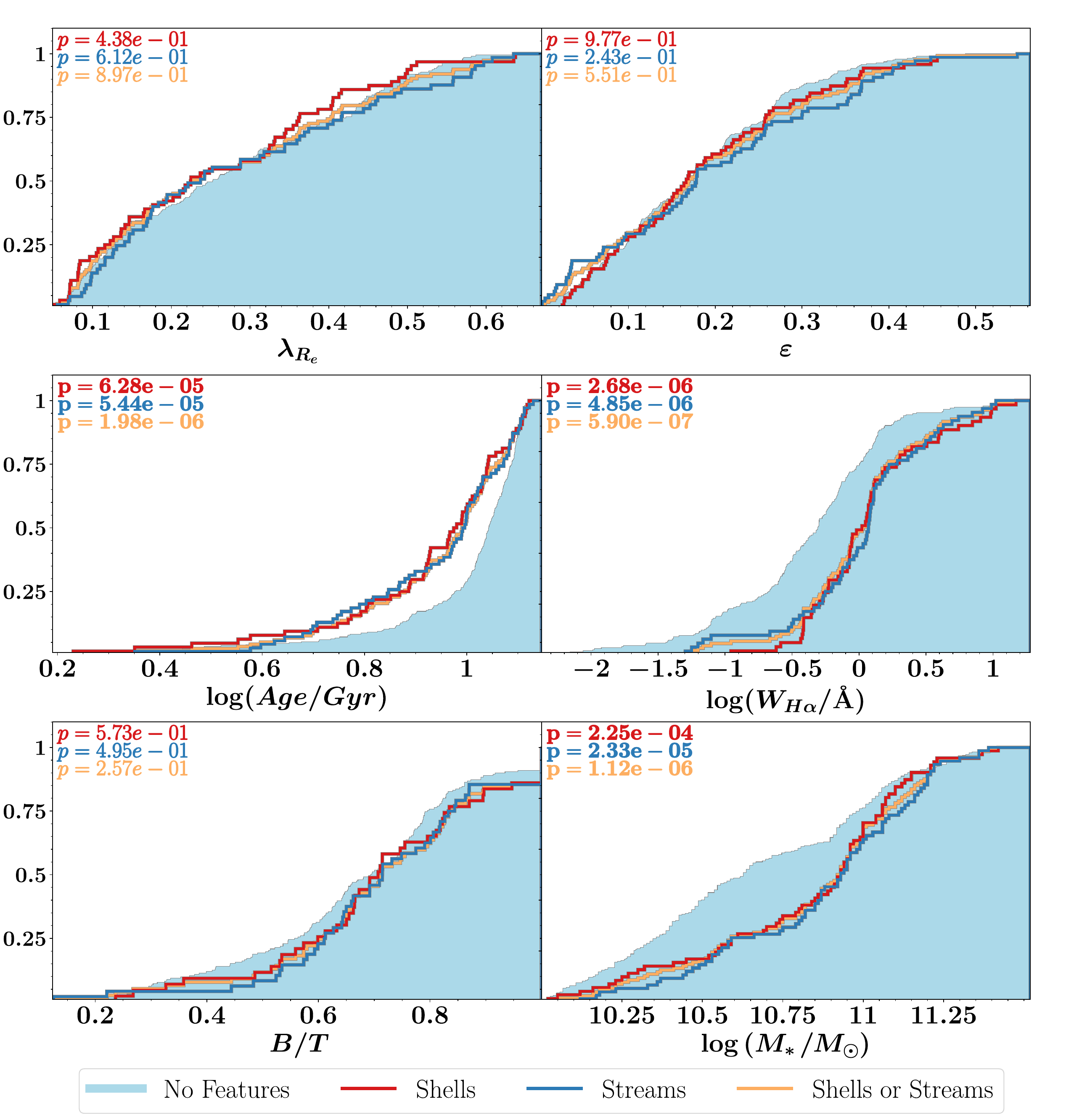}
    \caption{The cumulative distribution for all relevant parameters and tidal feature samples. Regular galaxies (sky blue), shells (red), streams (dark blue) and shells or streams (orange) are all shown. In each panel, we show different parameters. We show the p-values from a KS test between regular galaxies and shells (red), regular galaxies and streams (dark blue), and regular galaxies and combined features (orange). P-values less than $p=0.05$ are shown in bold. \update{We don't consider p-values above $p=0.05$ to be significant correlations}. We find that while features are correlated with lower stellar ages and higher W$_{\text{H}\upalpha}$, there is no such correlation with \lre\ or B/T.}
    \label{fig:dists}
\end{figure*}
\begin{figure*}
	\includegraphics[width=\textwidth]{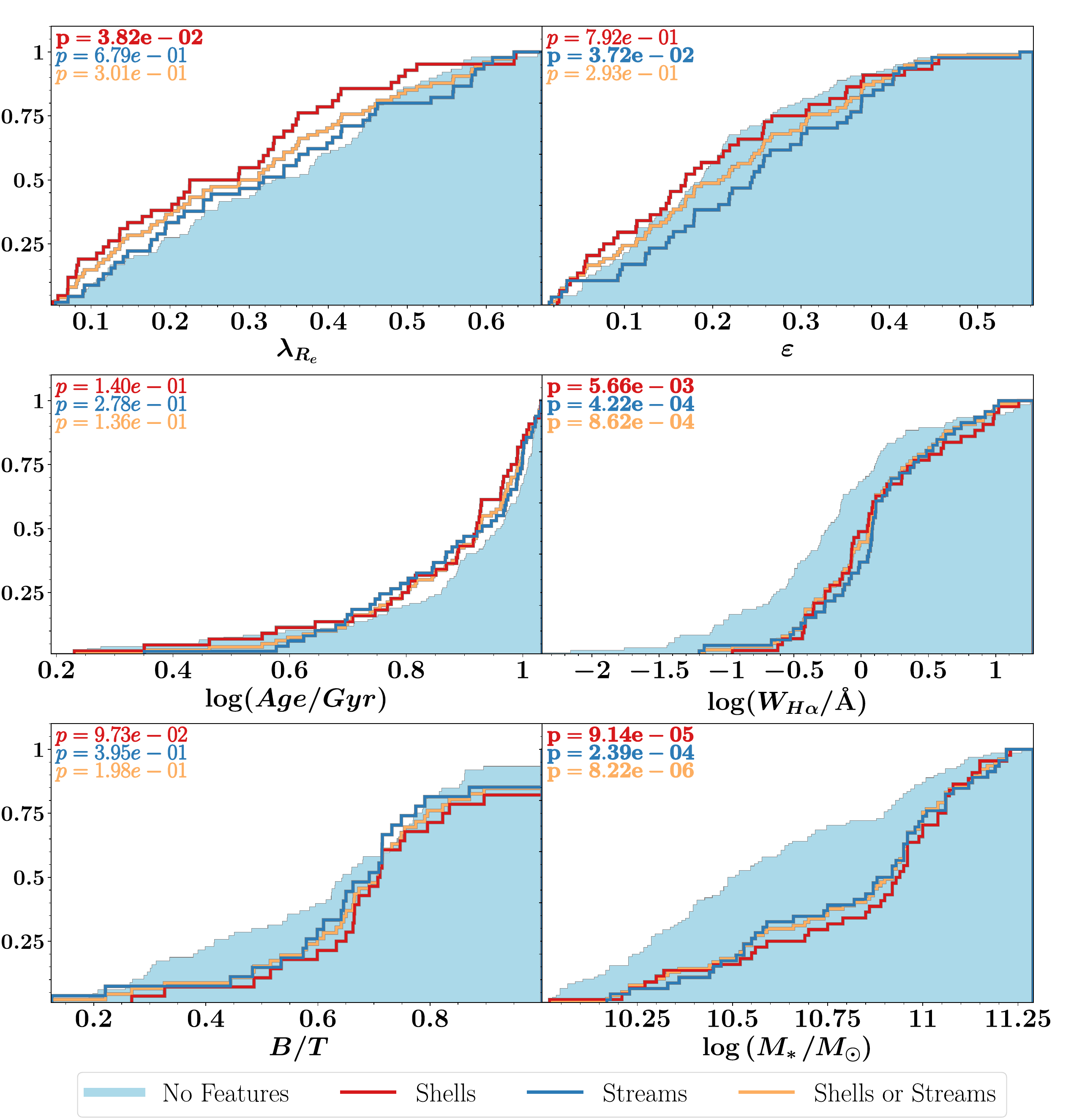}
    \caption{The cumulative distribution for all relevant parameters and tidal feature samples, for galaxies with stellar ages below the median light-weighted stellar age. All cumulative distributions and labels are the same as Figure \ref{fig:dists}. We find that for these galaxies with low stellar ages, shells are correlated with lower \lre and higher W$_{\text{H}\upalpha}$.}
    \label{fig:young_dists}
\end{figure*}
\begin{table*}
	\centering
	\caption{We present p-values from comparing distributions for an array of parameters for all galaxies and young galaxies, splitting our feature sample into shells and streams separated, and strong features only. A two sample KS test is applied to a set of feature galaxy parameters (e.g. shell galaxy \lre\ values) and the regular galaxy parameters (e.g. regular galaxy \lre\ values). Strong features are defined as features which were identified with a confidence level of at least 3/5 during classification. Significant p-values ($p < 0.05$) are highlighted in bold.}
	\label{tab:pvals}
	\begin{tabular}{ ccccccc } 
		\hline
		 & \multicolumn{3}{c}{Tidal Features} & \multicolumn{3}{c}{Strong Tidal Features}\\
		\hline
		\hline
		\textbf{All Galaxies} & Shells & Streams & Streams \& Shells & Shells & Streams & Shells \& Streams\\
		\hline
        \lre\ & $4.38\times 10^{-1}$ & $6.12\times 10^{-1}$ & $8.97\times 10^{-1}$ & $5.23\times 10^{-2}$ & $8.07\times 10^{-1}$ & $1.80\times 10^{-1}$\\
        $\varepsilon$ & $9.77\times 10^{-1}$ & $2.43\times 10^{-1}$ & $5.51\times 10^{-1}$ & $6.05\times 10^{-1}$ & $4.27\times 10^{-1}$ & $8.93\times 10^{-1}$\\
        $\log_{10}(\text{Age}/\text{Gyr})$ & $\bf{6.28\times 10^{-5}}$ & $\bf{5.44\times 10^{-5}}$ & $\bf{1.98\times 10^{-6}}$ & $\bf{2.08\times 10^{-2}}$ & $2.60\times 10^{-1}$ & $\bf{1.47\times 10^{-2}}$\\  
        $\log_{10}(\text{W}_{\text{H}\upalpha})$ & $\bf{2.68\times 10^{-6}}$ & $\bf{4.85\times 10^{-6}}$ & $\bf{5.90\times 10^{-7}}$ & $\bf{1.40\times 10^{-2}}$ & $\bf{3.26\times 10^{-2}}$ & $\bf{1.52\times 10^{-3}}$\\
        B$/$T & $5.73\times 10^{-1}$ & $4.95\times 10^{-1}$ & $2.57\times 10^{-1}$ & $7.44\times 10^{-1}$ & $2.96\times 10^{-1}$ & $4.92\times 10^{-1}$\\
        $\log_{10}(M_*/M_{\odot})$ & $\bf{2.25\times 10^{-4}}$ & $\bf{2.33\times 10^{-5}}$ & $\bf{1.12\times 10^{-6}}$ & $1.12\times 10^{-1}$ & $3.72\times 10^{-1}$ & $5.91\times 10^{-2}$\\
        \hline
		\textbf{Young Galaxies} & Shells & Streams & Streams \& Shells & Shells & Streams & Shells \& Streams\\
		\hline
        \lre\ & $\bf{3.82\times 10^{-2}}$ & $6.79\times 10^{-1}$ & $3.01\times 10^{-1}$ & $5.93\times 10^{-2}$ & $2.91\times 10^{-1}$ & $5.60\times 10^{-2}$\\
        $\varepsilon$ & $7.92\times 10^{-1}$ & $\bf{3.72\times 10^{-2}}$ & $2.93\times 10^{-1}$ & $9.95\times 10^{-1}$ & $7.59\times 10^{-1}$ & $9.34\times 10^{-1}$\\
        $\log_{10}(\text{Age}/\text{Gyr})$ & $1.40\times 10^{-1}$ & $2.78\times 10^{-1}$ & $1.36\times 10^{-1}$ & $\bf{3.90\times 10^{-2}}$ & $4.29\times 10^{-1}$ & $8.88\times 10^{-2}$\\
        $\log_{10}(\text{W}_{\text{H}\upalpha})$ & $\bf{5.66\times 10^{-3}}$ & $\bf{4.22\times 10^{-4}}$ & $\bf{8.62\times 10^{-4}}$ & $3.05\times 10^{-1}$ & $2.61\times 10^{-1}$ & $9.80\times 10^{-2}$\\
        B$/$T & $9.73\times 10^{-2}$ & $3.95\times 10^{-1}$ & $1.98\times 10^{-1}$ & $7.33\times 10^{-1}$ & $6.29\times 10^{-1}$ & $5.20\times 10^{-1}$\\
        $\log_{10}(M_*/M_{\odot})$ & $\bf{9.14\times 10^{-5}}$ & $\bf{2.39\times 10^{-4}}$ & $\bf{8.22\times 10^{-6}}$ & $1.11\times 10^{-1}$ & $4.25\times 10^{-1}$ & $1.95\times 10^{-1}$\\
        \hline
	\end{tabular}
\end{table*}
Here, we present the results of our analysis of the merger features of our SAMI galaxies in the GAMA field regions with $\logm > 10$. In total, we have 411 galaxies with classifications. 129 galaxies have a feature identified (71 shells, 76 streams and 18 with both). For simplicity, we refer to galaxies with shells as "Shell Galaxies", galaxies with streams as "Stream Galaxies", galaxies with a shell and/or stream as "Feature Galaxies", and galaxies with no shell or stream as "Regular Galaxies". Feature classifications are used alongside SAMI kinematic data to examine the link between merger features and slow rotators.

\updatetwo{We begin by presenting the distributions of our regular galaxies, shell galaxies and stream galaxies in several parameter spaces, using the parameters \lre, \logm, Age and B/T. This can be seen in Figure \ref{fig:all_parms}. Galaxies with no features are coloured in sky blue, galaxies with shells are coloured in red, galaxies with streams are coloured in dark blue, and galaxies with streams and shells are coloured in orange. Shell galaxies classified with a strength of at least 3 are circled in black, referred to as "strong" shells. In panel (b), we show the median stellar mass and age with dashed lines.} 

\updatetwo{We see evidence in panel (b) of galaxy shells being correlated with both age and mass, with neither being the sole driver of the relation. When splitting this panel into four quadrants by the median stellar mass and median age ($\logm = 10.75$ and Age/Gyr = 10.80 respectively, shown by dashed lines), we find the following shell feature fractions:}
\begin{itemize}
    \item \update{Low mass, high age galaxies: $9/90$, ${10.0}${\raisebox{0.5ex}{\tiny$^{+4.1}_{-2.3}$}}\%}
    \item \update{High mass, high age galaxies: $11/103$, ${10.7}${\raisebox{0.5ex}{\tiny$^{+3.8}_{-2.3}$}}\%}
    \item \update{Low mass, low age galaxies: $13/104$, ${12.5}${\raisebox{0.5ex}{\tiny$^{+4.0}_{-2.6}$}}\%}
    \item \update{High mass, low age galaxies: $31/89$, ${34.8}${\raisebox{0.5ex}{\tiny$^{+5.3}_{-4.7}$}}\%}
\end{itemize}

\updatetwo{There is a substantial number and fraction of galaxies with features at high mass and low age, with the fraction of features clearly affected by both age and mass.}

We also visually represent and quantify the difference between shell galaxies, stream galaxies, and regular galaxies using cumulative distributions and Kolmogorov–Smirnov (KS) tests. The cumulative distributions for all relevant parameters and feature samples can be seen in Figure \ref{fig:dists}. We analyse feature galaxies split into shell (red) and stream (dark blue) samples, and combined (orange). The resulting p-values from each KS test are given in Figure \ref{fig:dists} and Table \ref{tab:pvals}. Table \ref{tab:pvals} also includes p-values for when galaxy features are only considered for a classified strength greater than three. Each p-value comes from a two sample KS test applied to a set of feature galaxy parameters (e.g. shell galaxy \lre\ values) and the set of regular galaxy parameters (e.g. regular galaxy \lre\ values). Essentially, they quantify the probability that these merger features are not correlated with a given parameter. \update{If a p-value is higher than 0.05, we don't consider this to be a significant correlation.} \updatetwo{To test whether there are any significant differences in the tails of the distribution, we also employ the Anderson-Darling test. We find that all p-values from Figures \ref{fig:dists} and \ref{fig:young_dists} stay either significant or non-significant under this test, with the exception that the difference between the ellipticity ($\varepsilon$) of young stream and regular galaxies in Figure \ref{fig:young_dists} is no longer significant.}
\begin{figure}
	\includegraphics[width=\columnwidth]{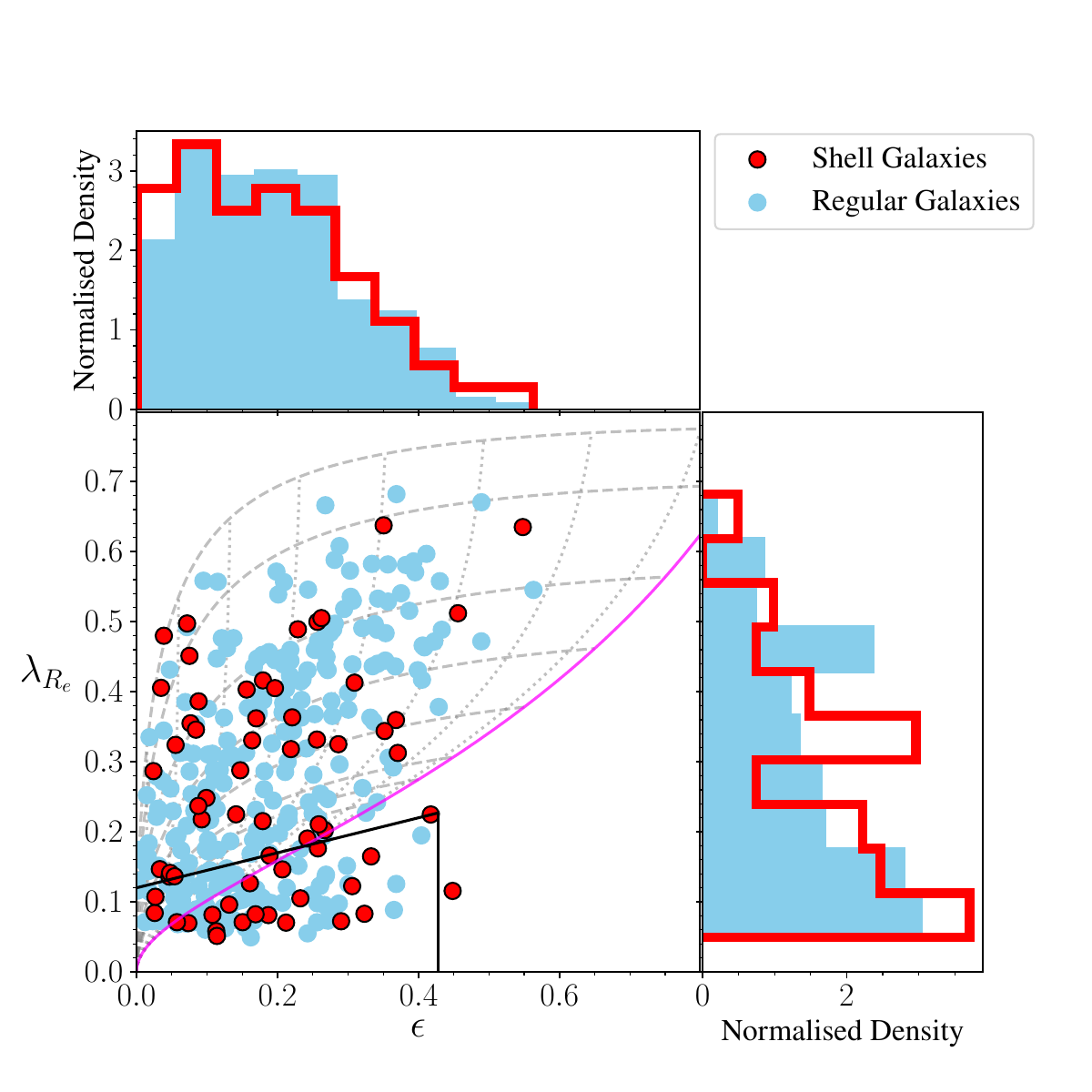}
    \caption{The distribution of selected SAMI galaxies in \lre-$\varepsilon$ space. Galaxies are coloured as red if they were identified as having shells, and by blue if they had no features identified. The solid magenta line shows the expected relation for edge-on oblate rotators with anisotropy $\beta_z = 0.7$. The dotted grey lines represent the same relation, but for varying inclinations ($10^{\circ}-80^{\circ}$). The dashed grey lines represent the same relation again, but for fixed ellipticity ($0.35-0.9$), changing with inclination. The black line denotes the separation between fast and slow rotators from \citet{2021MNRAS.505.3078V}, $0.12 + 0.25\varepsilon$ for $\varepsilon < 0.428$. Distributions of \lre\ and $\varepsilon$ are in the side panels, in red for galaxies with identified shells, and in blue for galaxies with no tidal features. The galaxies with shells do not have significantly different \lre\ values than galaxies without, and are not preferentially located in the slow rotator region.}
    \label{fig:spin_ellip_dist}
\end{figure}

\begin{figure}
	\includegraphics[width=\columnwidth]{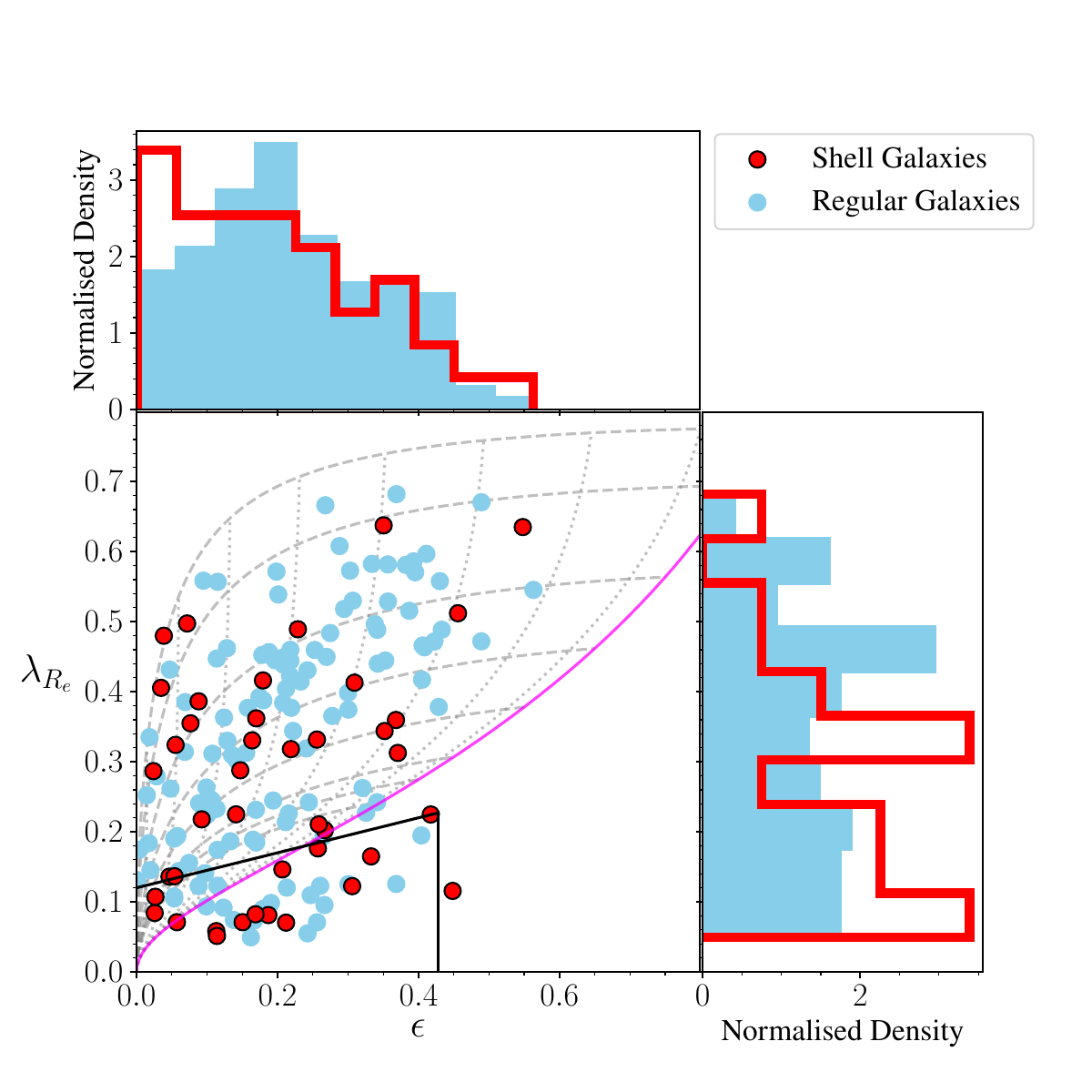}
    \caption{The distribution of selected SAMI galaxies with a mean stellar age below the median age in \lre-$\varepsilon$ space. The layout is similar to Figure \ref{fig:spin_ellip_dist}. The young galaxies with shells have significantly different \lre\ values than galaxies without, and are preferentially located in the slow rotator region.}
    \label{fig:spin_ellip_dist_young}
\end{figure}
We see in Figure \ref{fig:dists} that \lre, $\varepsilon$ and B/T are not correlated with any type of tidal feature. We additionally find that the mean stellar age is significantly lower, and H$\upalpha$ equivalent width (W$_{\text{H}\upalpha}$) \update{and stellar mass (\logm) are} significantly higher, for feature galaxies versus regular galaxies. Specifically, lower stellar age ($p = 6.28\times 10^{-5}$, $p=5.44\times 10^{-5}$), higher W$_{\text{H}\upalpha}$ ($p = 2.68\times 10^{-6}$, $p = 4.85 \times 10^{-6}$) \update{and higher \logm\  ($p = 2.25\times 10^{-4}, p = 2.35\times 10^{-5}$)} are correlated with both shells and streams. A lower mean stellar age is an expected result, as models and observations have shown that cold gas may be funnelled to the centre of galaxies during a merger, giving rise to central starburst activity and thus lower mean stellar ages \citep[e.g.,][]{2014MNRAS.437.2137S,2015MNRAS.454.1742K,2019MNRAS.482L..55T,2022MNRAS.514.3294B}. \update{An alternative interpretation is that features fade over time, so older galaxies may have had an earlier merger but have no identifiable features remaining. We also expect a higher stellar mass, given the correlation between stellar mass, number of mergers, and total flux in tidal features (i.e. identifiability of features) \citep[e.g.][]{2022MNRAS.513.1459M}.} We note here that we choose to analyse W$_{\text{H}\upalpha}$ instead of sSFR for our sample. We do this for several reasons. Firstly, as we have selected only ETGs, H$\upalpha$ based star formation rate measurements for these mostly passive galaxies are not very physically meaningful, and SED-fitted SFR values \citep{2022MNRAS.517.2677R} show no correlation. Secondly, W$_{\text{H}\upalpha}$ goes approximately with sSFR, without assumptions that H$\upalpha$ flux is from star formation sources.

\update{The correlation of higher stellar masses with tidal features as shown in Figures \ref{fig:dists} and \ref{fig:young_dists} is significant, however we found that it does not drive our strong mean stellar age result. In Figure \ref{fig:cum_dists_all_highmass}, we show that when restricting to galaxies with stellar masses above the median stellar mass ($\logm = 10.75$), there is no significant correlation between features and stellar mass. However, when we apply the same restriction, the qualitative results from Figure \ref{fig:spin_age} (see below) remain \updatetwo{(The mean spin values from this figure while only considering high-mass galaxies can additionally be seen in Table \ref{tab:spin_age_means})}. Specifically, we still see a lower spin for shell galaxies in the central age bin as compared to regular galaxies, and strong shell galaxies show the same relation with a larger gap in average spin.}

The stellar kinematics of galaxies are also well known to be affected by mergers \citep[e.g.,][]{2014MNRAS.444.3357N,2020IAUFM..30A.208L,2020MNRAS.493.3778S}. Given these results, an increase in B/T ratio and a decrease in $\varepsilon$ and \lre\ in feature galaxies is similarly expected, as slow rotators are described as high dispersion oblate spheroids. The lack of these results in Figure \ref{fig:dists}\updatetwo{, as well as the results from Figure \ref{fig:all_parms}} leads us to consider the role of stellar age, which we investigate in section \ref{sec:young_age_cum_hist}.

\subsection{Merger/Tidal Features in Galaxies with Relatively Young Age}
\label{sec:young_age_cum_hist}

We present our analysis of the galaxies in our sample with stellar ages below the median light-weighted stellar age (10.80 Gyr), henceforth referred to as younger galaxies. As tidal features are only detectable for $\sim 2-4$ Gyr \citep{2008MNRAS.391.1137L, 2010MNRAS.404..590L, 2010MNRAS.404..575L,2017MNRAS.465.2895L, 2021ApJ...912...45N, 2019A&A...632A.122M}, we expect that younger mean stellar age should be correlated with identifiable merger features. This can also be seen in Figure \ref{fig:dists}. The cumulative distributions for our young galaxies can be seen in Figure \ref{fig:young_dists}, presented similarly to Figure \ref{fig:dists}.

We find that when we restrict to lower ages, the presence of shells is correlated with lower \lre\ ($p = 0.038$), while the stellar age correlation disappears. This is likely due to reducing the dynamic range in stellar age. The correlation of lower W$_{\text{H}\upalpha}$ with features is maintained ($p = 0.0057$ for shells, $p = 0.00042$ for streams)\update{, as well as the correlation with higher \logm\  ($p = 9.14\times 10^{-5}$ for shells, $p = 2.39\times 10^{-4}$ for streams)}. The B/T ratio is not significantly correlated with features, however this is likely due to low number statistics. 268 of 411 galaxies have B/T ratios, of which only 77 ($\sim29\%$) have features. This can be compared to 325 galaxies with \lre\ measurements, of which 114 ($\sim 35\%$) have features. It should also be noted that \lre\ is measured within one effective radius, while B/T is a measurement taken from the full photometry of a galaxy, and thus the parameters are not indicative of the same physical scale.

From Figure \ref{fig:young_dists}, we see that even though the presence of shells is correlated with lower \lre, streams are not. Indeed, the analysis done to shell galaxies in Sections \ref{sec:spin_ellip_srs} and \ref{sec:role_of_mergers} below can be done to stream and combined feature galaxies as well, but all results remain strongly driven by shells. We also acknowledge that while streams may be more important than shells at lower stellar masses \citep[e.g.,][]{2023MNRAS.523.4381D}, our mass cut at $\logm = 10$ means our sample contains a reasonable fraction of shells. As a result of these considerations, this paper will henceforth be largely focused on shell features only.

\subsection{Spin, Ellipticity and Slow Rotators}
\label{sec:spin_ellip_srs}
When considering slow rotators, we look in more detail at the parameters that define a slow rotator, \lre\ and $\varepsilon$ (ellipticity). Figure \ref{fig:spin_ellip_dist} shows our galaxies with classifications in \lre-$\varepsilon$ space, where slow rotators are traditionally classified \citep[e.g.,][]{2011MNRAS.414..888E, 2016ARA&A..54..597C, 2021MNRAS.505.3078V}. We plot galaxies with shell features as red points, and galaxies without any type of feature as blue points. The magenta line represents the expected relation between \lre\ and ellipticity for edge-on oblate rotators with varying intrinsic ellipticity $\varepsilon_i$ and anisotropy $\beta_z = 0.7 \varepsilon_i$. Dotted grey lines show the same relation but for varying inclinations, with the grey dashed lines showing galaxies with fixed intrinsic ellipticity, changing with inclination. The black lines denote the separation between fast and slow rotators for seeing-corrected SAMI data, from \citet{2021MNRAS.505.3078V}. We find that feature galaxies are not preferentially located in the slow rotator region compared to regular galaxies ($32.8${\raisebox{0.5ex}{\tiny$^{+6.3}_{-5.3}$}}\% vs $30.7${\raisebox{0.5ex}{\tiny$^{+3.0}_{-2.7}$}}\% respectively)\footnote{Uncertainties are calculated as binomial confidence intervals from \cite{2011PASA...28..128C}.}. We note that the slow rotator fraction is particularly high here overall, due to a high mass, early-type sample. The distribution of younger galaxies in \lre-$\varepsilon$ space is shown in Figure \ref{fig:spin_ellip_dist_young}, with the same layout as Figure \ref{fig:spin_ellip_dist}. Shell galaxies in the younger sample are found to be significantly more preferentially located in the slow rotator region than regular younger galaxies ($31.0${\raisebox{0.5ex}{\tiny$^{+7.9}_{-6.1}$}}\% vs $17.39${\raisebox{0.5ex}{\tiny$^{+4.1}_{-3.0}$}}\% respectively). Note that the increase in significance when we cut by stellar age is due to the fraction of regular slow rotators falling, rather than the fraction of shell slow rotators rising.

\section{Discussion}
\label{sec:discussion}
In this section, we discuss the implications and limitations of our results and place them in context with similar studies.
\subsection{Effectiveness of Visual Classification}
\label{sec:visual_class}
In this work, visual classification of merger features using the combination of model subtraction with an interactive dynamic range in image cutouts provided a consistent method for identifying any apparent merger features. However, the completeness of our classifications is hampered by two factors. Firstly, the majority of tidal features are expected to lie at $\gtrsim 30$ mag arcsec$^{-2}$ \citep{2008ApJ...689..936J}, and our derived surface brightness limit for HSC $r$-band cutouts is $27.8 \pm 0.5$ mag arcsec$^{-2}$. Secondly, identification of tidal features at nominal surface brightness limits is significantly incomplete. Detectability is also strongly dependent on radial extent. For example, \citet{2018ApJ...866..103K} injected features into cutouts, with various surface brightnesses for each feature, and a derived surface brightness limit of $\sim 28$ mag arcsec$^{-2}$. They identified only $14.3\%$ of shells at $4R_e$ with a shell surface brightness of 27.0 mag arcsec$^{-2}$, as compared to a shell surface brightness of 25.125 mag arcsec$^{-2}$ which gave $64.3\%$ at $4R_e$ and $89.3\%$ at $5R_e$.  \citet{2022A&A...662A.124S} similarly were unable to find any tidal features fainter than 27.5 mag arcsec$^{-2}$, despite their nominal surface brightness limits of $28.3-29$ mag arcsec$^{-2}$.

Given these challenges in defining a completeness for our visual classifications, we can look to similar studies, examining the fraction of features as a comparison. The MATLAS survey \citep{2015MNRAS.446..120D,2020MNRAS.498.2138B,2020MNRAS.491.1901H,2022A&A...662A.124S} was examined by \citet{2020MNRAS.498.2138B} similarly to this work, classifying tidal features visually. \citet{2022arXiv220808443V} further investigated the correlation between tidal features and \lre\ for MATLAS, as well as a hydrodynamical cosmological simulation (\textit{Magneticum Pathfinder}\footnote{\url{www.magneticum.org}}, Dolag et al., in prep.). Comparing MATLAS to this work is appropriate, given the similar surface brightness limits ($28.3-29$ mag arcsec$^{-2}$ for MATLAS, $27.5\pm 0.5$ mag arcsec$^{-2}$ here, noting the discussion in Section \ref{sec:surf_bright} above concerning differences in calculated surface brightness limits) and target galaxies (field ETGs). The feature fractions from this work and from MATLAS can be seen in Table \ref{tab:feature_fractions}. The feature and shell fractions agree remarkably well between SAMI and MATLAS galaxies. Whilst the stream fractions are in slight disagreement, the overall results suggests that our work detects as many features as \citet{2020MNRAS.498.2138B}. Indeed, noting that MATLAS contains only galaxies in a local volume with distance below 42 Mpc, and SAMI contains galaxies out to a redshift of $\sim 0.1$ ($\sim 420$ Mpc), cosmological surface brightness dimming reduces intensities by a factor $(1+0.056)^4=1.246$ at the median redshift of our sample, and $(1+0.11)^4=1.509$ at the maximum redshift. This effectively reduces our surface brightness limit by 0.3 mag arcsec$^{-2}$ and 0.5 mag arcsec$^{-2}$ respectively.

\update{Additionally, other observations and simulations of samples like ours report similar fractions of features to this work. \cite{2020ApJS..247...43K} found a lower bound of $\sim 9.4\%$ of galaxies in local clusters showing shell features, and $\sim 22\%$ showing streams. \cite{2018MNRAS.480.1715P} found in the Illustris simulation that $18\pm 3 \%$ of massive, $z=0$ galaxies exhibited shells. Other studies \citep[e.g.][]{1988ApJ...328...88S,2009AJ....138.1417T} have found shells in between $\sim 10\%$ and $\sim 22\%$ of elliptical galaxies.}

\begin{table}
	\centering
	\caption{The number of galaxies with shells, streams and any feature for SAMI ($n_{\text{SAMI}}$) and MATLAS ($n_{\text{MATLAS}}$). We also show the fraction of features found as compared to the total number of galaxies for SAMI ($f_{\text{SAMI}}$) and MATLAS ($f_{\text{MATLAS}}$).}
	\label{tab:feature_fractions}
	\begin{tabular}{ ccccc } 
		\hline
		\hline
		 & Shells & Streams & Features & Total\\
		\hline
		$n_{\text{SAMI}}$ & $71$ & $76$ & $129$ & $411$\\
        $f_{\text{SAMI}}$ & $17.3${\raisebox{0.5ex}{\small$^{+2.0}_{-1.7}$}}\% & $18.5${\raisebox{0.5ex}{\small$^{+2.1}_{-1.8}$}}\% & $31.4${\raisebox{0.5ex}{\small$^{+2.4}_{-2.2}$}}\% & $--$\\
        $n_{\text{MATLAS}}$ & $23$ & $33$ & $41$ & $131$\\
        $f_{\text{MATLAS}}$ & $17.6${\raisebox{0.5ex}{\small$^{+3.8}_{-2.8}$}}\% & $25.2${\raisebox{0.5ex}{\small$^{+4.2}_{-3.4}$}}\% & $31.3${\raisebox{0.5ex}{\small$^{+4.3}_{-3.8}$}}\% & $--$\\
		\hline
		\hline
	\end{tabular}
\end{table}

Given the similar results and surface brightness limits between our work and MATLAS\update{, as well as similar feature fraction rates between our work and similar studies}, our combination of model subtraction with an interactive dynamic range in image cutouts likely provides a consistent and appropriate method of identifying tidal features around ETGs.

Instead of human visual classifications, machine learning techniques have recently shown promise in automatic classification \citep[e.g.,][]{2022MNRAS.511..100B,2022scio.confE...2B,2022MNRAS.514.3294B,2023MNRAS.521.3861D}, and make classifying very large datasets ($\gtrsim 10,000$ galaxies) possible in a reasonable amount of time. Whilst this is a method with a lot of potential, visual classification currently remains necessary for training these algorithms on real observations. However, developments in self-supervised learning could change this in the near future \citep[e.g.,][]{2023arXiv230704967D}.

The identification of LSB tidal features around galaxies is difficult, but future surveys which reach deeper surface brightness limits will be able to identify higher fractions of galaxies with features. \cite{2022MNRAS.513.1459M} showed that at a surface brightness limit of $\mu_r=35$, close to 100 percent of galaxies show some type of feature. Although a limiting surface brightness of 35 is still somewhat unrealistic, the importance of ranking the strength of tidal features as well as identification will become more important as they become more ubiquitous around galaxy images.
\subsection{Interpretation of H$\upalpha$ Equivalent Widths}
\begin{figure}
	\includegraphics[width=\columnwidth]{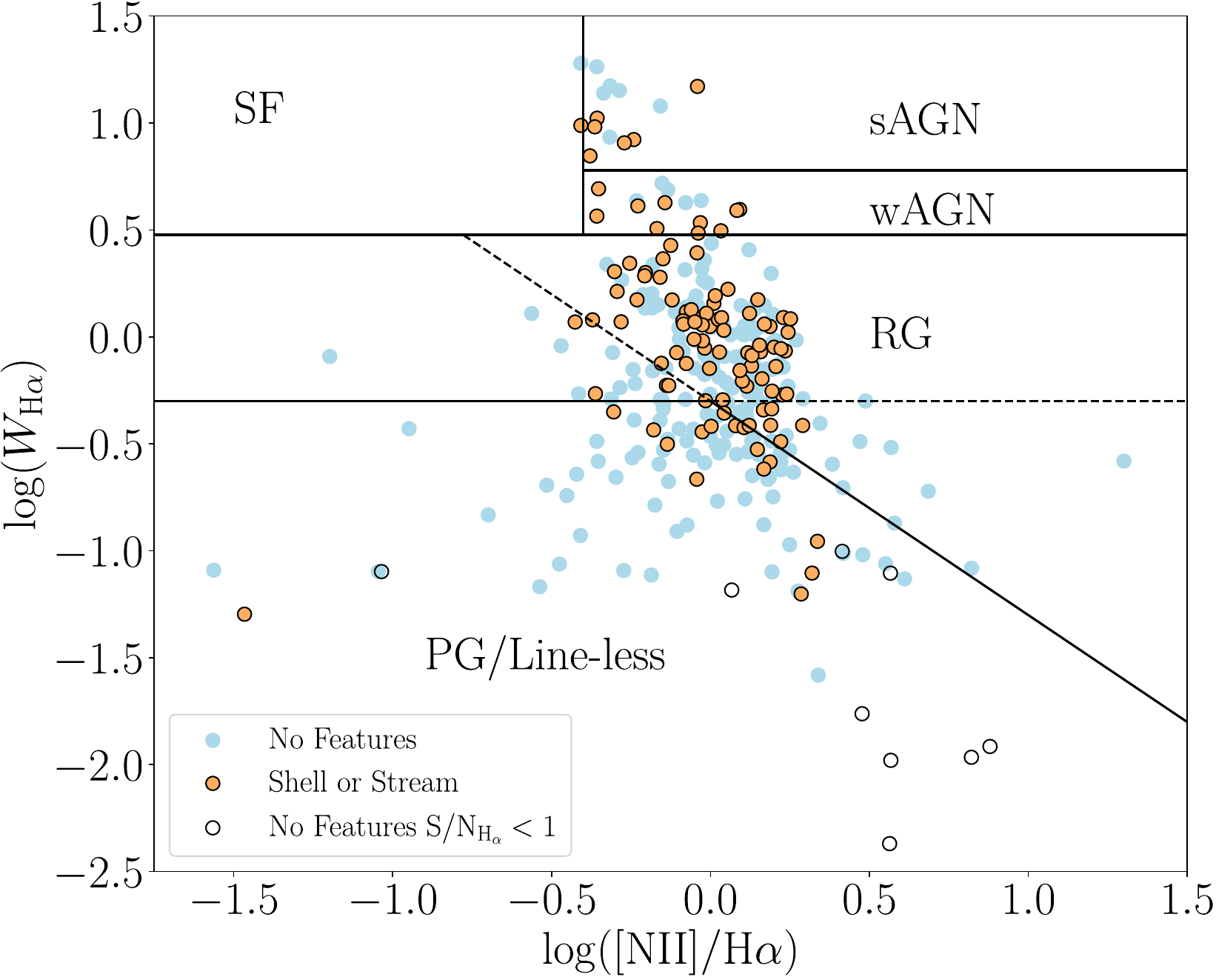}
    \caption{Distributions of our regular galaxies (blue) and feature galaxies (orange) in an emission line classification diagram, with [NII]/H$\upalpha$ line ratios plotted against W$_{\text{H}\upalpha}$. Galaxies with signal-to-noise ratios in W$_{\text{H}\upalpha}$ of below one are plotted as empty symbols, of which there are no feature galaxies. The following regions are defined: SF (star forming), sAGN (strong AGN), wAGN (weak AGN), RG (retired galaxies) and PG/Line-less (Passive galaxies/Line-less galaxes). The dashed lines denote areas that have one requirement for being a PG (W$_{\text{H}\upalpha}$ < 0.5 or W$_{\text{[NII]}}$ < 0.5). Feature galaxies are over-represented in the AGN regions, and under-represented in the PG region (see Table \ref{tab:cidfernandes_table}.}
    \label{fig:cidfernandes_bpt}
\end{figure}
\begin{table}
    \centering
    \caption{The number of galaxies within each region of Figure \ref{fig:cidfernandes_bpt}, separated into feature and regular galaxies. Percentages represent what fraction of each region comprises of regular and feature galaxies.}
    \label{tab:cidfernandes_table}
    \begin{tabular}{ cccc }
        \hline
        \hline
         & \textbf{Regular Galaxies} & \textbf{Feature Galaxies} & \textbf{Total}\\
         \hline
        \multirow{2}{3em}{\textbf{SF}} & 1 & 1 & 2\\
        & 50.0{\raisebox{0.5ex}{\small$^{+24.8}_{-24.8}$}}\% & 50.0{\raisebox{0.5ex}{\small$^{+24.8}_{-24.8}$}}\%\\
        \hline
        \multirow{2}{3em}{\textbf{sAGN}} & 6 & 6 & 12\\
        & 50.0{\raisebox{0.5ex}{\small$^{+13.4}_{-13.4}$}}\% & 50.0{\raisebox{0.5ex}{\small$^{+13.4}_{-13.4}$}}\%\\
        \hline
        \multirow{2}{3em}{\textbf{wAGN}} & 5 & 10 & 15\\
        & 33.3{\raisebox{0.5ex}{\small$^{+13.5}_{-9.6}$}}\% & 66.7{\raisebox{0.5ex}{\small$^{+9.6}_{-13.5}$}}\%\\
        \hline
        \multirow{2}{3em}{\textbf{RG}} & 122 & 72 & 194\\
        & 62.9{\raisebox{0.5ex}{\small$^{+3.3}_{-3.6}$}}\% & 37.1{\raisebox{0.5ex}{\small$^{+3.6}_{-3.3}$}}\%\\
        \hline
        \multirow{2}{3em}{\textbf{PG}} & 75 & 16 & 91\\
        & 82.4{\raisebox{0.5ex}{\small$^{+3.3}_{-4.7}$}}\% & 17.6{\raisebox{0.5ex}{\small$^{+4.7}_{-3.3}$}}\%\\
        \hline
        \multirow{2}{3em}{\textbf{All}} & 209 & 105 & 314\\
        & 66.6{\raisebox{0.5ex}{\small$^{+2.6}_{-2.8}$}}\% & 33.4{\raisebox{0.5ex}{\small$^{+2.8}_{-2.6}$}}\%\\
        \hline
    \end{tabular}
\end{table}
In Figures \ref{fig:dists} and \ref{fig:young_dists}, we showed that shell and stream galaxies have significantly higher H$\upalpha$ equivalent width (W$_{\text{H}\upalpha}$) values than regular galaxies. We are careful not to interpret larger amounts of H$\upalpha$ emission as indicative of star formation, given our sample was selected to be massive ETGs which are likely to be mostly passive. While a thorough investigation of the nature of H$\upalpha$ emission in these galaxies is beyond the scope of this paper, we use the classification method from \cite{2011MNRAS.413.1687C}, which utilises W$_{\text{H}\upalpha}$ and the [NII]/H$\upalpha$ line ratio to classify the emission source of weak-lined galaxies.

In Figure \ref{fig:cidfernandes_bpt} we show [NII]/H$\upalpha$ line ratio versus W$_{\text{H}\upalpha}$ values of regular (blue) and feature (orange) galaxies. Of the 314 galaxies with measured line ratios, 105 ($33\%$) are feature galaxies. Table \ref{tab:cidfernandes_table} shows the proportion of galaxies within each region of Figure \ref{fig:cidfernandes_bpt}. We note that galaxies with a W$_{\text{H}\upalpha}$ S/N < 1 are still included in our analysis (empty sumbols in Figure \ref{fig:cidfernandes_bpt}, because these low S/N values are typically driven by low W$_{\text{H}\upalpha}$ values, and removing them would heavily bias our sample.

We find that feature galaxies are over-represented in the sAGN and wAGN regions (50.0{\raisebox{0.5ex}{\small$^{+13.4}_{-13.4}$}}\% and 66.7{\raisebox{0.5ex}{\small$^{+9.6}_{-13.5}$}}\% vs 33.4{\raisebox{0.5ex}{\small$^{+2.8}_{-2.6}$}}\% overall), and under-represented in the passive galaxy regions (17.6{\raisebox{0.5ex}{\small$^{+4.7}_{-3.3}$}}\% vs 33.4{\raisebox{0.5ex}{\small$^{+2.8}_{-2.6}$}}\% overall). We also note that there are only 2 galaxies in the SF region, as expected given our ETG selection.

\cite{2018MNRAS.481.1774H} suggests that it is possible that the difference in emission between liny RGs (those with emission lines) and line-less RGs (PGs in this work) is due to warm gas content. They suggest RGs experienced a recent wet merger, providing gas and possibly inducing a period of star formation. This agrees broadly with our work, where we find feature galaxies under-represented in the PGs population. This is consistent with a scenario where ETGs that have recently experienced a merger (RGs) have increased their gas fraction, and thus show stronger emission lines from this gas than ETGs which have not experienced a recent merger (PGs).

\subsection{The Role of Galaxy Mergers in Kinematic Evolution and the Formation of Slow Rotators}
\label{sec:role_of_mergers}
Although slow rotators are an important subset of galaxy populations, their formation pathway remains unclear. Simulations have suggested that galaxy mergers are capable of providing the required (spin-down) morphological transformation \citep[e.g.,][]{2009A&A...501L...9D, 2009MNRAS.397.1202J,2011MNRAS.416.1654B,2014MNRAS.444.3357N,2017ApJ...837...68C,2017MNRAS.464.3850L,2020IAUFM..30A.208L,2018MNRAS.476.4327L,2017MNRAS.468.3883P,2020MNRAS.493.3778S}, but conclusive observational evidence linking tidal features to slow rotators is yet to be found. Indeed, recent studies \citep[e.g.,][]{2022MNRAS.509.4372L} suggest that there is a diverse number of formation pathways for slow rotator galaxies.

\begin{figure*}
	\includegraphics[width=\textwidth]{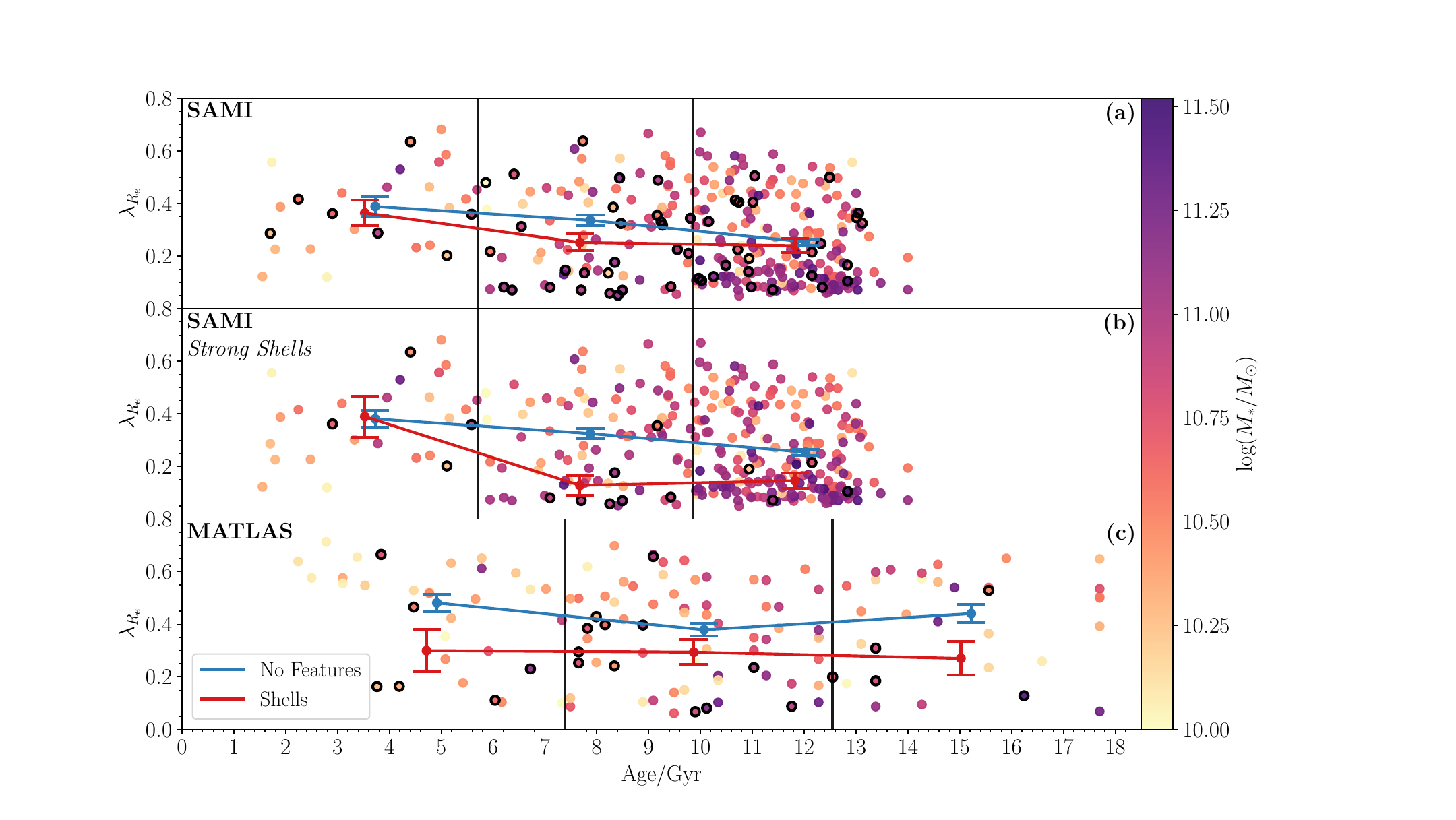}
    \caption{The distribution of SAMI and MATLAS galaxies in \lre-mean light-weighted stellar age space. Galaxies are coloured by \logm, and galaxies with shells are circled in black. The average \lre\ value in 3 age bins is drawn for regular galaxies in blue, and for shell galaxies in red. Panel (a) shows SAMI galaxies, panel (b) shows SAMI galaxies with only strong shells considered (strength classified as 3/5 or higher), and panel (c) shows MATLAS galaxies. The age bins are taken to be equally spaced between the minimum and maximum stellar ages. As MATLAS stellar ages are derived differently to SAMI's and have significant differences (such as different maximum ages), we don't compare them numerically, and thus using the same age bins is not necessary. We see that in the central age bin, SAMI galaxies have a correlation between shells and lower \lre. This correlation grows stronger when only considering strong shells.}
    \label{fig:spin_age}
\end{figure*}

\begin{figure*}
	\includegraphics[width=\textwidth]{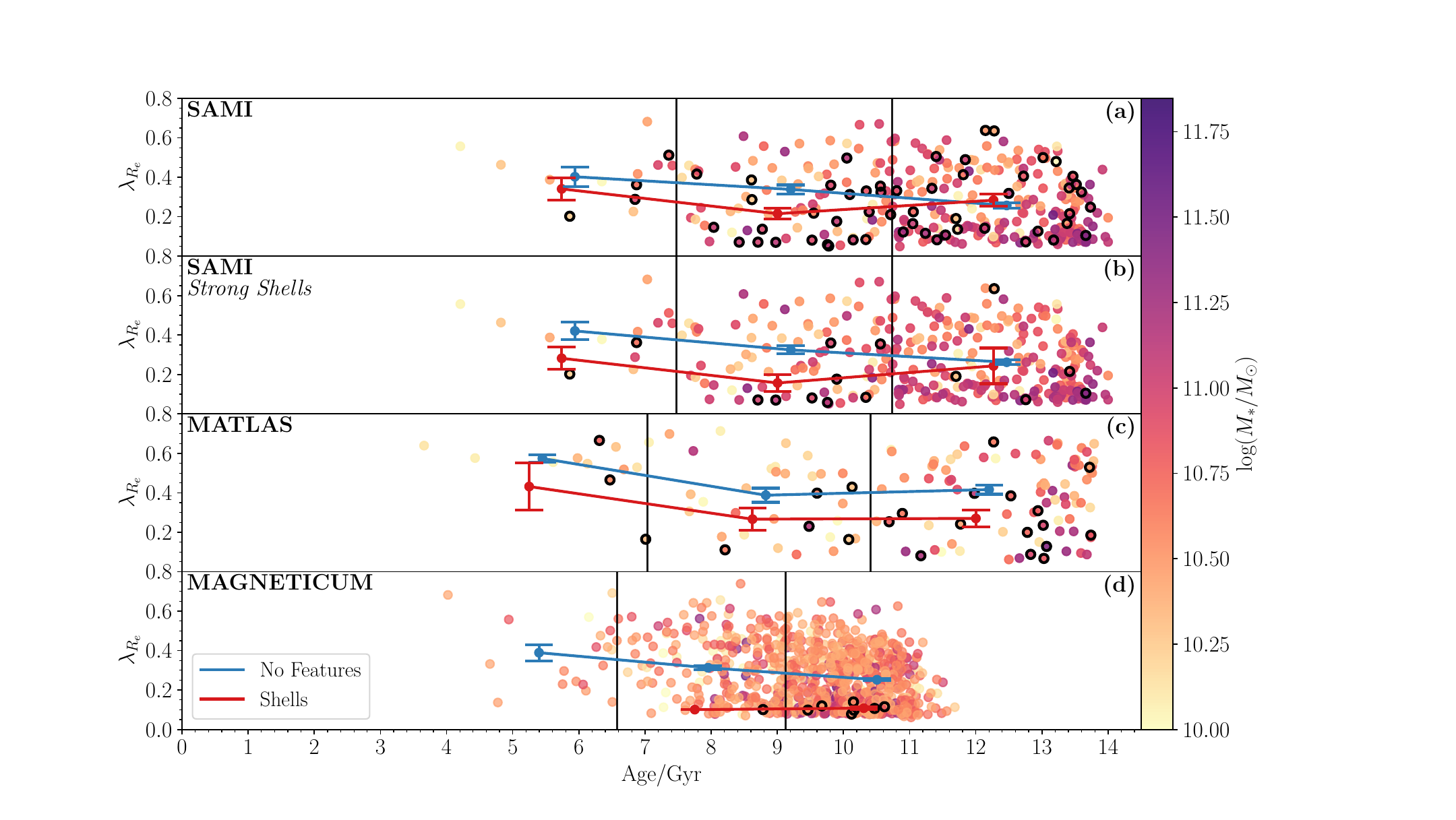}
    \caption{The distribution of SAMI, MATLAS and Magneticum galaxies in \lre-mean mass-weighted stellar age space. Galaxies are coloured by \logm, and galaxies with shells are circled in black. The average \lre\ value in 3 age bins is drawn for regular galaxies in blue, and for shell galaxies in red. Panel (a) shows SAMI galaxies, panel (b) shows SAMI galaxies with only strong shells considered (strength classified as 3/5 or higher), panel (c) shows MATLAS galaxies, and panel (d) shows Magneticum galaxies. Mass-weighted stellar ages show a typical lack of ages below $\sim 6$ Gyr as compared to the light-weighted ages (seen in Figure \ref{fig:spin_age}), due to light-weighted averages favouring young, bright stars.}
    \label{fig:spin_age_mass_weighted}
\end{figure*}
\begin{table*}
	\centering
	\caption{The mean spin (\lre) within each of the three light-weighted age bins in Figure \ref{fig:spin_age}, for panels (a), (b) and (c). \updatetwo{We also include the values for an equivalent plot to Figure \ref{fig:spin_age}, but only considering galaxies with $\logm \geq 10.75$.}}
	\label{tab:spin_age_means}
	\begin{tabular}{ cccccc } 
		\hline
		\hline
		 & & Regular Galaxies & Shell Galaxies & Massive Regular Galaxies & Massive Shell Galaxies\\
      & & \lre\ Values & \lre\ Values & \lre\ Values & \lre\ Values\\
		\hline
		\hline
		 & Lowest Age Bin & $0.39\pm 0.04$ & $0.36 \pm 0.05$ & $0.35 \pm 0.07$ & $0.24 \pm 0.06$\\
		 SAMI (a) & Middle Age Bin & $0.34\pm 0.02$ & $0.25 \pm 0.03$ & $0.31 \pm 0.03$ & $0.21 \pm 0.03$\\
		 & Highest Age Bin & $0.25\pm 0.01$ & $0.24\pm 0.03$ & $0.21 \pm 0.02$ & $0.23 \pm 0.04$\\
		 \hline
		 & Lowest Age Bin & $0.38\pm 0.03$ & $0.39\pm 0.08$ & $0.31 \pm 0.05$ & $0.22 \pm 0.09$\\
		 SAMI Strong Shells (b) & Middle Age Bin & $0.33\pm 0.02$ & $0.13\pm 0.04$ & $0.29 \pm 0.02$ & $0.14 \pm 0.09$\\
		 & Highest Age Bin & $0.25 \pm 0.01$ & $0.15\pm 0.03$ & $0.21 \pm 0.01$ & $0.09 \pm 0.01$\\
		 \hline
		 & Lowest Age Bin & $0.48\pm 0.03$ & $0.30\pm 0.08$ & $0.51 \pm 0.07$ & $0.38 \pm 0.09$\\
		 MATLAS (c) & Middle Age Bin & $0.38\pm 0.02$ & $0.29\pm 0.05$ & $0.36 \pm 0.04$ & $0.17 \pm 0.06$\\
		 & Highest Age Bin & $0.44\pm 0.03$ & $0.27\pm 0.06$ & $0.39 \pm 0.08$ & $0.21 \pm 0.04$\\
		\hline
		\hline
	\end{tabular}
\end{table*}

We find that for early type galaxies with relatively young stellar ages of $\lesssim 10$Gyr, the presence of a shell is correlated with lower \lre. We further show this relationship in Figure \ref{fig:spin_age}, with mean values in Table \ref{tab:spin_age_means}. In this figure, we show SAMI and MATLAS galaxies above $\logm > 10$ in \lre-Light Weighted Age space. Galaxies are coloured by stellar mass, and galaxies with identified shells are circled in black. Further, each sample is divided into three equally spaced age bins. The average spin in each bin for both galaxies with no features, and galaxies with shells, is plotted for each bin. Panel (a) contains SAMI galaxies, panel (b) is the same as panel (a) but with the adjustment that only \textit{strong} shells are considered (classified strength of at least 3/5), and panel (c) contains MATLAS galaxies.

The age values shown in Figure \ref{fig:spin_age} have different physical interpretations for SAMI and MATLAS. Stellar ages for SAMI are derived from full spectral fitting to SSP models, with a maximum stellar age of 14 Gyr. The age measurement is derived as a light-weighted mean within 1$R_e$. Age measurements for MATLAS are derived in \citet{2015MNRAS.448.3484M}, from SSP models which used H$\beta$, Fe5015, Mg$b$ and Fe5270 line indices. These SSP ages \update{also} differ from SAMI's in that they go to ages above the currently accepted age of the universe \citep{2016A&A...594A..13P}. As a result, we treat them as relative ages, which we will not compare numerically to the SAMI ages. We choose to focus on light-weighted ages here as they may be a better proxy for the time since the last major merger than mass-weighted ages.

We first see the overall trend in Figure \ref{fig:spin_age} that spin decreases towards higher ages in SAMI galaxies (\citet{2018NatAs...2..483V}, Croom et. al in prep). However, we also see a separation in the mean spin of shell and regular galaxies in the central age bin for all panels.

The results in Figure \ref{fig:spin_age}, panel (a), suggest that for SAMI galaxies in the lowest age bin, there is no difference in \lre\ for galaxies that have a shell and those that do not. There is an indirect link between stellar age and time since merger, as mergers can induce star formation and a light-weighted mean stellar age approximately measures the time since the last period of star formation. This is not a direct correlation however, as dry mergers for example may not induce star formation, and may result in a galaxy with a mean stellar age older than the time since the dry merger. We suggest that feature galaxies in the lowest stellar age bin may have only recently merged. At this stage, the galaxy is still disk-like and has not yet undergone the spin-down morphological transformation. Alternatively, galaxies in the lowest age bin may also have merged longer ago, but experienced a wet merger which re-built a star-forming disc. These two possible scenarios both result in shell galaxies with no difference in \lre\ to regular galaxies. A further possibility is simply that a combination of low number statistics and noise results in no significant difference between shell galaxies and regular galaxies in the low age bin.

For the central age bin, we see significantly lower spin for galaxies with shells than galaxies with no features. This is an expected relationship between merger features and kinematics from simulations \citep[e.g.,][]{2009A&A...501L...9D,2009MNRAS.397.1202J,2011MNRAS.416.1654B,2014MNRAS.444.3357N,2017ApJ...837...68C,2017MNRAS.468.3883P,2017MNRAS.464.3850L,2018MNRAS.473.4956L,2020MNRAS.493.3778S,2020IAUFM..30A.208L}, which suggest that galaxy mergers spin down galaxies. 

Finally, in the oldest age bin, we see no difference in the spin of galaxies with shells and those without. This could be due to several reasons. Firstly, some of the oldest galaxies may have simply been born hot, with any mergers having no effect on spin. Additionally, it could plausibly be because the spin-down merger occurred long enough ago ($\gtrsim 4$ Gyr) that the shell features are no longer identifiable. In these cases, the correlation between shells and spin disappears. 

The scenario of fading tidal features is further supported by the transition of shell and regular galaxies from the central to the oldest age bin in panel (a) in Figure \ref{fig:spin_age}. While the shell galaxies do not change their spin, the regular galaxies have lower spin at older ages. This would be consistent with a contamination of regular galaxies with merger galaxies that no longer have identifiable shells. Further, Figures \ref{fig:spin_ellip_dist} and \ref{fig:spin_ellip_dist_young} show that when only considering younger galaxies, feature galaxies are preferentially located in the slow rotator region of a \lre-$\varepsilon$ diagram. This is not the case when considering all galaxies, regardless of age. The difference between this younger galaxy result and the result when considering all galaxies is driven not by fewer shell galaxy slow rotators at older ages, but more regular galaxy slow rotators at older ages. This is again consistent with merger-induced slow rotators no longer having identifiable shells at long enough times post-merger.

We address the possibility of weak shells in actuality being rings or weak spiral arms in panel (b) of Figure \ref{fig:spin_age}. By selecting only shells identified with an average strength of at least 3/5, we have much higher confidence that we are not selecting any fast rotating discs without any real features. The strength measurement both correlates with higher confidence in a feature being real, and the visual strength of the feature. We see, that in comparison to panel (a), the difference in mean spins in the middle age bin is much more significant, and a significant difference arises in the highest age bin. This is likely due to almost all of the high spin shell galaxies in the central and highest age bin only having low strengths, reducing the impact of fading shells on this metric.

\update{An alternative to the shell features fading in the oldest galaxies is that the mergers that generate features add mass with a lower mean stellar age. Given the median ages and masses for our shell and regular samples, we can estimate the mass ratio of mergers required to form a shell galaxy from a regular galaxy.} 

\update{If we assume that a galaxy from the regular population undergoes a merger event(s), it needs enough mass added to match the median shell galaxy stellar mass. In order to match the lower stellar age of shell galaxies, we need a mass-weighted stellar age estimate for a galaxy of a given mass. We fit a linear relation between stellar mass and mass-weighted stellar age for all SAMI ETGs, shown in Equation \ref{eq:age_mass_fit}:}

\begin{equation}
    \label{eq:age_mass_fit}
    \text{Age/Gyr} = 2.10\times \logm -10.86
\end{equation}

\update{We note here that there is considerable scatter in this relation, and this approach only provides a simplified assessment of the types of mergers required. We can calculate the mass-weighted stellar age for a galaxy after it has experienced a series of mergers. We find that, given our results, a galaxy needs to undergo several $\sim 1:4-6$ mergers to transform from a regular galaxy to a shell galaxy. Two equal mass mergers will not work, as the galaxies will typically have similar age. However, it's worth bearing in mind that we do expect features to fade over time, and this may be the main reason for not seeing shells in the oldest galaxies. If this is the case, high-redshift 1:1 mergers could form the oldest galaxies without features. }

\subsection{Comparison of SAMI, MATLAS and Magneticum}
\label{sec:sami_matlas_comparison}
Due to MATLAS light-weighted ages being derived signficantly differently to SAMI light-weighted ages, we do not make direct numerical comparisons between these ages. Rather, we use $\textit{mass}$-weighted SAMI ages from full spectral fitting, as mass-weighted ages derived in a similar way are available for MATLAS. This gives the additional advantage that we can compare to the simulated Magneticum galaxies which were also analysed by \citet{2022arXiv220808443V}, as they have mass-weighted age measurements available. We show this comparison in Figure \ref{fig:spin_age_mass_weighted}, with a similar layout to Figure \ref{fig:spin_age}. The Magneticum galaxies used in this work come from Box4 (uhr), a ($48 \text{Mpc}/h$)$^3$ box, initially containing $2\times 576^3$ particles of  DM (dark matter) and gas. The particles have masses of $m_{\text{DM}} = 3.6\times 10^7M_{\odot}/h$ and $m_{\text{gas}}=7.3\times 10^6M_{\odot}/h$, with a gravitational softening length of 1.4 kpc$/h$ for DM and gas, and 0.7 kpc$/h$ for star particles. More detail on the Magneticum simulations can be found in \cite{2015ApJ...812...29T}.

MATLAS shows no difference in spin in the lowest age bin, and a lower spin for shell galaxies in the central and oldest age bin in panel (c), Figure \ref{fig:spin_age_mass_weighted}, consistent with our SAMI results in the central bin. The main disagreement between SAMI and MATLAS is in the oldest age bin, where SAMI does not show a difference in spin. There are several possible reasons for this. Firstly, linear age measurements for galaxies, particularly at old ages, have an inherently large scatter. The oldest age bin may not be particularly meaningful physically as an age measurement. Secondly, LSB features in SAMI are more affected by cosmological redshift surface brightness dimming than MATLAS (see Section \ref{sec:visual_class}). We expect that galaxies in the oldest age bin are more likely to have experienced the longest time since their latest merger, and thus their features are expected to be fainter, and more susceptible to becoming undetectable through redshift dimming. 

Magneticum galaxies ages are represented by mass-weighted mean stellar lookback time within 1$R_{\text{half}}$, and shown in panel (d) of Figure \ref{fig:spin_age_mass_weighted}. Whilst SAMI and MATLAS galaxies are sorted by visual morphology to exclusively examine early types, Magneticum defines morphology in terms of their kinematics, and as such correlating to their kinematics after a kinematics based cut is not appropriate. We attempt to account for this by performing a cut between ETGs and LTGs on the star forming main sequence of Magneticum, in comparison to SAMI (see Figure \ref{fig:magneticum_etg_cut}). We find that Magneticum shell galaxies have exclusively low spin values compared to regular galaxies in both the central and oldest age bins, with no detected shell galaxies in the lowest age bin. While Magneticum galaxies do trend towards lower spin with age as SAMI does, the lack of any scatter in the spin of shell galaxies is clearly different from our SAMI and MATLAS results. Whilst it is possible that this is caused by low number statistics, it is also possible that Magneticum is not fully capturing the possibility of a galaxy rebuilding a star-forming disc post radial infall merger.

\subsection{Radial Infall Galaxy Mergers}
\label{sec:radial_infall_mergers}

Our results show that there is an excess of shells around younger slow rotator galaxies. We suggest that radial path mergers play an important role in the formation of the slow rotator population. Given the diverse formation pathways of slow rotators \citep{2022MNRAS.509.4372L}, the existence of shells around a slow rotator is a potential method to distinguish this formation scenario from others. However, galaxies which had these mergers early have shells which are very difficult to detect with current methods and surveys. These features are likely to be undetectable after $\sim 4$ Gyr even with improved surveys \citep{2019A&A...632A.122M, 2022MNRAS.513.1459M}. A more significant result can be found by only considering "strong" shells, but the number of detections drops significantly (71 total shells vs 19 strong shells in this work).

These merger scenarios would also result in an excess of shells around galaxies with a higher B/T ratio, as mergers are known to impact the growth of the bulge component of galaxies \citep[e.g.,][]{2012MNRAS.423.1544S, 2013MNRAS.433.2986W, 2022MNRAS.516.3569B}. However, we do not see a correlation of B/T with shell galaxies. In Figure \ref{fig:all_parms}, we examined this by showing the distributions of our galaxies in all relevant parameter spaces, using the parameters \lre, \logm, Age and B/T. 

In panel (e), although all shell galaxies are spread throughout the \lre-B/T distribution, strong shell galaxies are much more tightly clustered in a high B/T, low \lre\ region. Similarly in panel (d), strong shell galaxies are mostly clustered in a high \logm region. Low \lre, high B/T and high stellar mass define a slow rotator. Indeed, ${52.9}${\raisebox{0.5ex}{\tiny$^{+11.1}_{-11.8}$}}\% (9/17) of strong shell galaxies are slow rotators (cf. $32.8${\raisebox{0.5ex}{\tiny$^{+6.3}_{-5.3}$}}\%  (21/64) for all shell galaxies).

There is a possible correlation between B/T, \lre\ and shells, but it is complex, and the strength of the shells as well as stellar age plays an important role. Although the strength represents the classifiers' confidence of a shell, it also correlates with the surface brightness of the feature, and thus likely increases with the mass of the most recent merger. 

\section{Conclusions}
\label{sec:conclusion}

We investigated the role of mergers in the formation of slow rotator galaxies through the identification of low surface brightness tidal features in deep HSC imaging. We performed a visual inspection of HSC images, with surface brightness model subtracted residuals and galaxy cutouts with an interactive dynamic range to classify SAMI ETGs with $\logm >10$ as either galaxies with shells, galaxies with streams, or galaxies with no tidal features. We further connected these features to stellar kinematics and morphology to investigate a link between galaxy mergers and the removal of angular momentum in galaxies.

We find that shell features are correlated with lower \lre\ in early-type galaxies with stellar ages below the median (Age/Gyr = 10.8), with a p-value of $p=3.82\times 10^{-2}$. Further, the signal is strongest at intermediate ages ($7.39<\text{Age/Gyr}<12.55$), with the average \lre\ for galaxies with shell features being significantly lower than for galaxies with no features. \update{Additionally, although our results show that feature fraction depends on both age and stellar mass, this result is not driven by stellar mass. At high stellar masses, our relation between \lre, age and shells remains, and the fraction of galaxies with shells has a strong dependence on age (${34.8}${\raisebox{0.5ex}{\tiny$^{+5.3}_{-4.7}$}}\% for low age, high mass galaxies; ${10.7}${\raisebox{0.5ex}{\tiny$^{+3.8}_{-2.3}$}}\% for high age, high mass galaxies).}

Galaxies with obvious, high surface brightness shells are correlated with lower \lre\ in all but the youngest mean stellar age galaxies. These galaxies generally have low \lre, high B/T, high stellar mass, and are more likely to be slow rotators (${52.9}${\raisebox{0.5ex}{\tiny$^{+11.1}_{-11.8}$}}\% (9/17) for strong shells vs ${31.1}${\raisebox{0.5ex}{\tiny$^{+2.7}_{-2.5}$}}\% (101/325) for our full sample).

Radial galaxy merger events (i.e. those with small impact parameter) likely play an important role in the formation of the slow rotator population. These radial mergers additionally produces shell-like tidal features, which can further be used to distinguish this formation pathway for slow rotators from others. The relation between radial mergers and angular momentum is complex, with our results showing that shell features, shell feature strength, \lre, stellar age, stellar mass and B/T ratio all play a role.

We find that lower mean stellar age ($p=1.98\times 10^{-6}$) and higher H$\upalpha$ EW ($p=3.32\times10^{-6}$) are correlated with tidal features in early-type galaxies. The emission line properties of feature galaxies are consistent with a scenario in which ETGs experience a merger, increasing both their gas content and emission line strength. 

Future surveys \citep[such as the LSST, ][]{2019ApJ...873..111I,2020arXiv200111067B,2022MNRAS.513.1459M} which reach even fainter surface brightness limits, as well as more advanced feature identification through potential methods such as machine learning, will allow for higher completeness in feature identification, and the possibility of orders of magnitude more galaxies inspected. A more detailed and quantitative analysis of the impact of mergers on slow rotator evolution will then be possible. Additionally, methods such as dynamical modelling of orbital populations in slow rotators will provide an alternate method to identify merger histories via the modelled orbital populations. Finally, the in-progress Hector Survey will be able to explore the environmental effects on mergers \citep{2016SPIE.9908E..1FB}, enabling a similar analysis as performed here in diverse environments such as the outskirts of clusters.

\section*{Acknowledgements}
The SAMI Galaxy Survey is based on observations made at the Anglo-Australian Telescope. SAMI was developed jointly by the University of Sydney and the Australian Astronomical Observatory (AAO). The SAMI input catalogue is based on data taken from the Sloan Digital Sky Survey, the GAMA Survey and the VST ATLAS Survey. The SAMI Galaxy Survey is supported by the Australian Research Council (ARC) Centre of Excellence ASTRO 3D (CE170100013) and CAASTRO (CE110001020), and other participating institutions. The SAMI Galaxy Survey website is \url{http://samisurvey.org/}. 

The calculations for the Magneticum Box4 hydrodynamical simulation were carried out at the Leibniz Supercomputer Center (LRZ) under the project pr83li (Magneticum).

The SAMI instrument was funded by the AAO and JBH through FF0776384, LE130100198. JBH is supported by an ARC Laureate Fellowship and an ARC Federation Fellowship that funded the SAMI prototype. FDE acknowledges funding through the H2020 ERC Consolidator Grant 683184. SO acknowledges support from the NRF grant funded by the Korea government (MSIT) (No. 2020R1A2C3003769 and No. RS-2023-00214057). LMV acknowledges support by the German Academic Scholarship Foundation (Studienstiftung des deutschen Volkes) and the Marianne-Plehn-Program from the Elite Network of Bavaria. JvdS acknowledges support of an Australian Research Council Discovery Early Career Research Award (project number DE200100461) funded by the Australian Government. JJB acknowledges support of an Australian Research Council Future Fellowship (FT180100231). AR acknowledges the receipt of a Scholarship for International Research Fees (SIRF) and an International Living Allowance Scholarship (Ad Hoc Postgraduate Scholarship) at The University of Western Australia.

The Hyper Suprime-Cam (HSC) collaboration includes the astronomical communities of Japan and Taiwan, and Princeton University. The HSC instrumentation and software were developed by the National Astronomical Observatory of Japan (NAOJ), the Kavli Institute for the Physics and Mathematics of the Universe (Kavli IPMU), the University of Tokyo, the High Energy Accelerator Research Organization (KEK), the Academia Sinica Institute for Astronomy and Astrophysics in Taiwan (ASIAA), and Princeton University. Funding was contributed by the FIRST program from the Japanese Cabinet Office, the Ministry of Education, Culture, Sports, Science and Technology (MEXT), the Japan Society for the Promotion of Science (JSPS), Japan Science and Technology Agency (JST), the Toray Science Foundation, NAOJ, Kavli IPMU, KEK, ASIAA, and Princeton University. 

This paper makes use of software developed for Vera C. Rubin Observatory. We thank the Rubin Observatory for making their code available as free software at http://pipelines.lsst.io/.

This paper is based on data collected at the Subaru Telescope and retrieved from the HSC data archive system, which is operated by the Subaru Telescope and Astronomy Data Center (ADC) at NAOJ. Data analysis was in part carried out with the cooperation of Center for Computational Astrophysics (CfCA), NAOJ. We are honored and grateful for the opportunity of observing the Universe from Maunakea, which has the cultural, historical and natural significance in Hawaii. 

The Pan-STARRS1 Surveys (PS1) and the PS1 public science archive have been made possible through contributions by the Institute for Astronomy, the University of Hawaii, the Pan-STARRS Project Office, the Max Planck Society and its participating institutes, the Max Planck Institute for Astronomy, Heidelberg, and the Max Planck Institute for Extraterrestrial Physics, Garching, The Johns Hopkins University, Durham University, the University of Edinburgh, the Queen’s University Belfast, the Harvard-Smithsonian Center for Astrophysics, the Las Cumbres Observatory Global Telescope Network Incorporated, the National Central University of Taiwan, the Space Telescope Science Institute, the National Aeronautics and Space Administration under grant No. NNX08AR22G issued through the Planetary Science Division of the NASA Science Mission Directorate, the National Science Foundation grant No. AST-1238877, the University of Maryland, Eotvos Lorand University (ELTE), the Los Alamos National Laboratory, and the Gordon and Betty Moore Foundation.

\section*{Data Availability}
The results of the data displayed in the figures can be obtained by contacting the corresponding author upon request. The images originate from the Hyper-Suprime Cam Public Data Release 2 \citep[HSC-PDR2;][]{2019PASJ...71..114A}  and are accessible at \url{https://hsc-release.mtk.nao.ac.jp/doc/index.php/tools-2/}. The data utilized in this paper, including kinematic measurements, originate from SAMI Data Release 3 \citep{2021MNRAS.505..991C}. This data is accessible through Australian Astronomical Optics’ Data Central at \url{https://datacentral.org.au/}.



\bibliographystyle{mnras}
\bibliography{ads_references} 

\begin{thebibliography}{}
\makeatletter
\relax
\def\mn@urlcharsother{\let\do\@makeother \do\$\do\&\do\#\do\^\do\_\do\%\do\~}
\def\mn@doi{\begingroup\mn@urlcharsother \@ifnextchar [ {\mn@doi@}
  {\mn@doi@[]}}
\def\mn@doi@[#1]#2{\def\@tempa{#1}\ifx\@tempa\@empty \href
  {http://dx.doi.org/#2} {doi:#2}\else \href {http://dx.doi.org/#2} {#1}\fi
  \endgroup}
\def\mn@eprint#1#2{\mn@eprint@#1:#2::\@nil}
\def\mn@eprint@arXiv#1{\href {http://arxiv.org/abs/#1} {{\tt arXiv:#1}}}
\def\mn@eprint@dblp#1{\href {http://dblp.uni-trier.de/rec/bibtex/#1.xml}
  {dblp:#1}}
\def\mn@eprint@#1:#2:#3:#4\@nil{\def\@tempa {#1}\def\@tempb {#2}\def\@tempc
  {#3}\ifx \@tempc \@empty \let \@tempc \@tempb \let \@tempb \@tempa \fi \ifx
  \@tempb \@empty \def\@tempb {arXiv}\fi \@ifundefined
  {mn@eprint@\@tempb}{\@tempb:\@tempc}{\expandafter \expandafter \csname
  mn@eprint@\@tempb\endcsname \expandafter{\@tempc}}}

\bibitem[\protect\citeauthoryear{{Ahn} et~al.,}{{Ahn}
  et~al.}{2012}]{2012ApJS..203...21A}
{Ahn} C.~P.,  et~al., 2012, \mn@doi [\apjs] {10.1088/0067-0049/203/2/21}, \href
  {https://ui.adsabs.harvard.edu/abs/2012ApJS..203...21A} {203, 21}

\bibitem[\protect\citeauthoryear{{Aihara} et~al.,}{{Aihara}
  et~al.}{2018}]{2018PASJ...70S...4A}
{Aihara} H.,  et~al., 2018, \mn@doi [\pasj] {10.1093/pasj/psx066}, \href
  {https://ui.adsabs.harvard.edu/abs/2018PASJ...70S...4A} {70, S4}

\bibitem[\protect\citeauthoryear{{Aihara} et~al.,}{{Aihara}
  et~al.}{2019}]{2019PASJ...71..114A}
{Aihara} H.,  et~al., 2019, \mn@doi [\pasj] {10.1093/pasj/psz103}, \href
  {https://ui.adsabs.harvard.edu/abs/2019PASJ...71..114A} {71, 114}

\bibitem[\protect\citeauthoryear{{Allen} et~al.,}{{Allen}
  et~al.}{2015}]{2015MNRAS.446.1567A}
{Allen} J.~T.,  et~al., 2015, \mn@doi [\mnras] {10.1093/mnras/stu2057}, \href
  {https://ui.adsabs.harvard.edu/abs/2015MNRAS.446.1567A} {446, 1567}

\bibitem[\protect\citeauthoryear{{Atkinson}, {Abraham}  \&
  {Ferguson}}{{Atkinson} et~al.}{2013}]{2013ApJ...765...28A}
{Atkinson} A.~M.,  {Abraham} R.~G.,   {Ferguson} A. M.~N.,  2013, \mn@doi
  [\apj] {10.1088/0004-637X/765/1/28}, \href
  {https://ui.adsabs.harvard.edu/abs/2013ApJ...765...28A} {765, 28}

\bibitem[\protect\citeauthoryear{{Barsanti} et~al.,}{{Barsanti}
  et~al.}{2022}]{2022MNRAS.516.3569B}
{Barsanti} S.,  et~al., 2022, \mn@doi [\mnras] {10.1093/mnras/stac2405}, \href
  {https://ui.adsabs.harvard.edu/abs/2022MNRAS.516.3569B} {516, 3569}

\bibitem[\protect\citeauthoryear{{Bell} et~al.,}{{Bell}
  et~al.}{2006}]{2006ApJ...640..241B}
{Bell} E.~F.,  et~al., 2006, \mn@doi [\apj] {10.1086/499931}, \href
  {https://ui.adsabs.harvard.edu/abs/2006ApJ...640..241B} {640, 241}

\bibitem[\protect\citeauthoryear{{Bertin} \& {Arnouts}}{{Bertin} \&
  {Arnouts}}{1996}]{1996A&AS..117..393B}
{Bertin} E.,  {Arnouts} S.,  1996, \mn@doi [\aaps] {10.1051/aas:1996164}, \href
  {https://ui.adsabs.harvard.edu/abs/1996A&AS..117..393B} {117, 393}

\bibitem[\protect\citeauthoryear{{Bezanson} et~al.,}{{Bezanson}
  et~al.}{2018}]{2018ApJ...858...60B}
{Bezanson} R.,  et~al., 2018, \mn@doi [\apj] {10.3847/1538-4357/aabc55}, \href
  {https://ui.adsabs.harvard.edu/abs/2018ApJ...858...60B} {858, 60}

\bibitem[\protect\citeauthoryear{{Bickley}, {Ellison}, {Patton}, {Bottrell},
  {Gwyn}  \& {Hudson}}{{Bickley} et~al.}{2022}]{2022MNRAS.514.3294B}
{Bickley} R.~W.,  {Ellison} S.~L.,  {Patton} D.~R.,  {Bottrell} C.,  {Gwyn} S.,
    {Hudson} M.~J.,  2022, \mn@doi [\mnras] {10.1093/mnras/stac1500}, \href
  {https://ui.adsabs.harvard.edu/abs/2022MNRAS.514.3294B} {514, 3294}

\bibitem[\protect\citeauthoryear{{B{\'\i}lek} et~al.,}{{B{\'\i}lek}
  et~al.}{2020}]{2020MNRAS.498.2138B}
{B{\'\i}lek} M.,  et~al., 2020, \mn@doi [\mnras] {10.1093/mnras/staa2248},
  \href {https://ui.adsabs.harvard.edu/abs/2020MNRAS.498.2138B} {498, 2138}

\bibitem[\protect\citeauthoryear{{Bland-Hawthorn} et~al.,}{{Bland-Hawthorn}
  et~al.}{2011}]{2011OExpr..19.2649B}
{Bland-Hawthorn} J.,  et~al., 2011, \mn@doi [Optics Express]
  {10.1364/OE.19.002649}, \href
  {https://ui.adsabs.harvard.edu/abs/2011OExpr..19.2649B} {19, 2649}

\bibitem[\protect\citeauthoryear{{Bois} et~al.,}{{Bois}
  et~al.}{2011}]{2011MNRAS.416.1654B}
{Bois} M.,  et~al., 2011, \mn@doi [\mnras] {10.1111/j.1365-2966.2011.19113.x},
  \href {https://ui.adsabs.harvard.edu/abs/2011MNRAS.416.1654B} {416, 1654}

\bibitem[\protect\citeauthoryear{{Bottrell}}{{Bottrell}}{2022}]{2022scio.confE...2B}
{Bottrell} C.,  2022, in SciOps 2022: Artificial Intelligence for Science and
  Operations in Astronomy (SCIOPS). Proceedings of the ESA/ESO SCOPS Workshop
  held 16-20 May. p.~2, \mn@doi{10.5281/zenodo.6551859}

\bibitem[\protect\citeauthoryear{{Bottrell}, {Hani}, {Teimoorinia}, {Patton}
  \& {Ellison}}{{Bottrell} et~al.}{2022}]{2022MNRAS.511..100B}
{Bottrell} C.,  {Hani} M.~H.,  {Teimoorinia} H.,  {Patton} D.~R.,   {Ellison}
  S.~L.,  2022, \mn@doi [\mnras] {10.1093/mnras/stab3717}, \href
  {https://ui.adsabs.harvard.edu/abs/2022MNRAS.511..100B} {511, 100}

\bibitem[\protect\citeauthoryear{{Brough} et~al.,}{{Brough}
  et~al.}{2020}]{2020arXiv200111067B}
{Brough} S.,  et~al., 2020, \mn@doi [arXiv e-prints]
  {10.48550/arXiv.2001.11067}, \href
  {https://ui.adsabs.harvard.edu/abs/2020arXiv200111067B} {p. arXiv:2001.11067}

\bibitem[\protect\citeauthoryear{{Bryant}, {Bland-Hawthorn}, {Fogarty},
  {Lawrence}  \& {Croom}}{{Bryant} et~al.}{2014}]{2014MNRAS.438..869B}
{Bryant} J.~J.,  {Bland-Hawthorn} J.,  {Fogarty} L.~M.~R.,  {Lawrence} J.~S.,
  {Croom} S.~M.,  2014, \mn@doi [\mnras] {10.1093/mnras/stt2254}, \href
  {https://ui.adsabs.harvard.edu/abs/2014MNRAS.438..869B} {438, 869}

\bibitem[\protect\citeauthoryear{{Bryant} et~al.,}{{Bryant}
  et~al.}{2015}]{2015MNRAS.447.2857B}
{Bryant} J.~J.,  et~al., 2015, \mn@doi [\mnras] {10.1093/mnras/stu2635}, \href
  {https://ui.adsabs.harvard.edu/abs/2015MNRAS.447.2857B} {447, 2857}

\bibitem[\protect\citeauthoryear{{Bryant} et~al.,}{{Bryant}
  et~al.}{2016}]{2016SPIE.9908E..1FB}
{Bryant} J.~J.,  et~al., 2016, in {Evans} C.~J.,  {Simard} L.,   {Takami} H.,
  eds,  Society of Photo-Optical Instrumentation Engineers (SPIE) Conference
  Series Vol. 9908, Ground-based and Airborne Instrumentation for Astronomy VI.
  p. 99081F (\mn@eprint {arXiv} {1608.03921}), \mn@doi{10.1117/12.2230740}

\bibitem[\protect\citeauthoryear{{Byrd} \& {Howard}}{{Byrd} \&
  {Howard}}{1992}]{1992AJ....103.1089B}
{Byrd} G.~G.,  {Howard} S.,  1992, \mn@doi [\aj] {10.1086/116128}, \href
  {https://ui.adsabs.harvard.edu/abs/1992AJ....103.1089B} {103, 1089}

\bibitem[\protect\citeauthoryear{{Cameron}}{{Cameron}}{2011}]{2011PASA...28..128C}
{Cameron} E.,  2011, \mn@doi [\pasa] {10.1071/AS10046}, \href
  {https://ui.adsabs.harvard.edu/abs/2011PASA...28..128C} {28, 128}

\bibitem[\protect\citeauthoryear{{Cappellari}}{{Cappellari}}{2002}]{2002MNRAS.333..400C}
{Cappellari} M.,  2002, \mn@doi [\mnras] {10.1046/j.1365-8711.2002.05412.x},
  \href {https://ui.adsabs.harvard.edu/abs/2002MNRAS.333..400C} {333, 400}

\bibitem[\protect\citeauthoryear{{Cappellari}}{{Cappellari}}{2016}]{2016ARA&A..54..597C}
{Cappellari} M.,  2016, \mn@doi [\araa] {10.1146/annurev-astro-082214-122432},
  \href {https://ui.adsabs.harvard.edu/abs/2016ARA&A..54..597C} {54, 597}

\bibitem[\protect\citeauthoryear{{Cappellari}}{{Cappellari}}{2017}]{2017MNRAS.466..798C}
{Cappellari} M.,  2017, \mn@doi [\mnras] {10.1093/mnras/stw3020}, \href
  {https://ui.adsabs.harvard.edu/abs/2017MNRAS.466..798C} {466, 798}

\bibitem[\protect\citeauthoryear{{Cappellari} \& {Emsellem}}{{Cappellari} \&
  {Emsellem}}{2004}]{2004PASP..116..138C}
{Cappellari} M.,  {Emsellem} E.,  2004, \mn@doi [\pasp] {10.1086/381875}, \href
  {https://ui.adsabs.harvard.edu/abs/2004PASP..116..138C} {116, 138}

\bibitem[\protect\citeauthoryear{{Casura} et~al.,}{{Casura}
  et~al.}{2022}]{2022MNRAS.516..942C}
{Casura} S.,  et~al., 2022, \mn@doi [\mnras] {10.1093/mnras/stac2267}, \href
  {https://ui.adsabs.harvard.edu/abs/2022MNRAS.516..942C} {516, 942}

\bibitem[\protect\citeauthoryear{{Chabrier}}{{Chabrier}}{2003}]{2003PASP..115..763C}
{Chabrier} G.,  2003, \mn@doi [\pasp] {10.1086/376392}, \href
  {https://ui.adsabs.harvard.edu/abs/2003PASP..115..763C} {115, 763}

\bibitem[\protect\citeauthoryear{{Choi} \& {Yi}}{{Choi} \&
  {Yi}}{2017}]{2017ApJ...837...68C}
{Choi} H.,  {Yi} S.~K.,  2017, \mn@doi [\apj] {10.3847/1538-4357/aa5e4b}, \href
  {https://ui.adsabs.harvard.edu/abs/2017ApJ...837...68C} {837, 68}

\bibitem[\protect\citeauthoryear{{Cid Fernandes}, {Stasi{\'n}ska}, {Mateus}  \&
  {Vale Asari}}{{Cid Fernandes} et~al.}{2011}]{2011MNRAS.413.1687C}
{Cid Fernandes} R.,  {Stasi{\'n}ska} G.,  {Mateus} A.,   {Vale Asari} N.,
  2011, \mn@doi [\mnras] {10.1111/j.1365-2966.2011.18244.x}, \href
  {https://ui.adsabs.harvard.edu/abs/2011MNRAS.413.1687C} {413, 1687}

\bibitem[\protect\citeauthoryear{{Cortese} et~al.,}{{Cortese}
  et~al.}{2016}]{2016MNRAS.463..170C}
{Cortese} L.,  et~al., 2016, \mn@doi [\mnras] {10.1093/mnras/stw1891}, \href
  {https://ui.adsabs.harvard.edu/abs/2016MNRAS.463..170C} {463, 170}

\bibitem[\protect\citeauthoryear{{Croom} et~al.,}{{Croom}
  et~al.}{2012}]{2012MNRAS.421..872C}
{Croom} S.~M.,  et~al., 2012, \mn@doi [\mnras]
  {10.1111/j.1365-2966.2011.20365.x}, \href
  {https://ui.adsabs.harvard.edu/abs/2012MNRAS.421..872C} {421, 872}

\bibitem[\protect\citeauthoryear{{Croom} et~al.,}{{Croom}
  et~al.}{2021}]{2021MNRAS.505..991C}
{Croom} S.~M.,  et~al., 2021, \mn@doi [\mnras] {10.1093/mnras/stab229}, \href
  {https://ui.adsabs.harvard.edu/abs/2021MNRAS.505..991C} {505, 991}

\bibitem[\protect\citeauthoryear{{D'Eugenio} et~al.,}{{D'Eugenio}
  et~al.}{2021}]{2021MNRAS.504.5098D}
{D'Eugenio} F.,  et~al., 2021, \mn@doi [\mnras] {10.1093/mnras/stab1146}, \href
  {https://ui.adsabs.harvard.edu/abs/2021MNRAS.504.5098D} {504, 5098}

\bibitem[\protect\citeauthoryear{{Desmons}, {Brough}  \& {Lanusse}}{{Desmons}
  et~al.}{2023a}]{2023arXiv230704967D}
{Desmons} A.,  {Brough} S.,   {Lanusse} F.,  2023a, \mn@doi [arXiv e-prints]
  {10.48550/arXiv.2307.04967}, \href
  {https://ui.adsabs.harvard.edu/abs/2023arXiv230704967D} {p. arXiv:2307.04967}

\bibitem[\protect\citeauthoryear{{Desmons}, {Brough}, {Mart{\'\i}nez-Lombilla},
  {De Propris}, {Holwerda}  \& {L{\'o}pez-S{\'a}nchez}}{{Desmons}
  et~al.}{2023b}]{2023MNRAS.523.4381D}
{Desmons} A.,  {Brough} S.,  {Mart{\'\i}nez-Lombilla} C.,  {De Propris} R.,
  {Holwerda} B.,   {L{\'o}pez-S{\'a}nchez} {\'A}.~R.,  2023b, \mn@doi [\mnras]
  {10.1093/mnras/stad1639}, \href
  {https://ui.adsabs.harvard.edu/abs/2023MNRAS.523.4381D} {523, 4381}

\bibitem[\protect\citeauthoryear{{Di Matteo}, {Jog}, {Lehnert}, {Combes}  \&
  {Semelin}}{{Di Matteo} et~al.}{2009}]{2009A&A...501L...9D}
{Di Matteo} P.,  {Jog} C.~J.,  {Lehnert} M.~D.,  {Combes} F.,   {Semelin} B.,
  2009, \mn@doi [\aap] {10.1051/0004-6361/200912354}, \href
  {https://ui.adsabs.harvard.edu/abs/2009A&A...501L...9D} {501, L9}

\bibitem[\protect\citeauthoryear{{Dom{\'\i}nguez S{\'a}nchez}
  et~al.,}{{Dom{\'\i}nguez S{\'a}nchez} et~al.}{2023}]{2023MNRAS.521.3861D}
{Dom{\'\i}nguez S{\'a}nchez} H.,  et~al., 2023, \mn@doi [\mnras]
  {10.1093/mnras/stad750}, \href
  {https://ui.adsabs.harvard.edu/abs/2023MNRAS.521.3861D} {521, 3861}

\bibitem[\protect\citeauthoryear{{Driver} et~al.,}{{Driver}
  et~al.}{2011}]{2011MNRAS.413..971D}
{Driver} S.~P.,  et~al., 2011, \mn@doi [\mnras]
  {10.1111/j.1365-2966.2010.18188.x}, \href
  {https://ui.adsabs.harvard.edu/abs/2011MNRAS.413..971D} {413, 971}

\bibitem[\protect\citeauthoryear{{Duc} et~al.,}{{Duc}
  et~al.}{2015}]{2015MNRAS.446..120D}
{Duc} P.-A.,  et~al., 2015, \mn@doi [\mnras] {10.1093/mnras/stu2019}, \href
  {https://ui.adsabs.harvard.edu/abs/2015MNRAS.446..120D} {446, 120}

\bibitem[\protect\citeauthoryear{{Emsellem} et~al.,}{{Emsellem}
  et~al.}{2007}]{2007MNRAS.379..401E}
{Emsellem} E.,  et~al., 2007, \mn@doi [\mnras]
  {10.1111/j.1365-2966.2007.11752.x}, \href
  {https://ui.adsabs.harvard.edu/abs/2007MNRAS.379..401E} {379, 401}

\bibitem[\protect\citeauthoryear{{Emsellem} et~al.,}{{Emsellem}
  et~al.}{2011}]{2011MNRAS.414..888E}
{Emsellem} E.,  et~al., 2011, \mn@doi [\mnras]
  {10.1111/j.1365-2966.2011.18496.x}, \href
  {https://ui.adsabs.harvard.edu/abs/2011MNRAS.414..888E} {414, 888}

\bibitem[\protect\citeauthoryear{{Fraser-McKelvie} et~al.,}{{Fraser-McKelvie}
  et~al.}{2021}]{2021MNRAS.503.4992F}
{Fraser-McKelvie} A.,  et~al., 2021, \mn@doi [\mnras] {10.1093/mnras/stab573},
  \href {https://ui.adsabs.harvard.edu/abs/2021MNRAS.503.4992F} {503, 4992}

\bibitem[\protect\citeauthoryear{{Green} et~al.,}{{Green}
  et~al.}{2018}]{2018MNRAS.475..716G}
{Green} A.~W.,  et~al., 2018, \mn@doi [\mnras] {10.1093/mnras/stx3135}, \href
  {https://ui.adsabs.harvard.edu/abs/2018MNRAS.475..716G} {475, 716}

\bibitem[\protect\citeauthoryear{{Habas} et~al.,}{{Habas}
  et~al.}{2020}]{2020MNRAS.491.1901H}
{Habas} R.,  et~al., 2020, \mn@doi [\mnras] {10.1093/mnras/stz3045}, \href
  {https://ui.adsabs.harvard.edu/abs/2020MNRAS.491.1901H} {491, 1901}

\bibitem[\protect\citeauthoryear{{Harborne}, {van de Sande}, {Cortese},
  {Power}, {Robotham}, {Lagos}  \& {Croom}}{{Harborne}
  et~al.}{2020}]{2020MNRAS.497.2018H}
{Harborne} K.~E.,  {van de Sande} J.,  {Cortese} L.,  {Power} C.,  {Robotham}
  A.~S.~G.,  {Lagos} C.~D.~P.,   {Croom} S.,  2020, \mn@doi [\mnras]
  {10.1093/mnras/staa1847}, \href
  {https://ui.adsabs.harvard.edu/abs/2020MNRAS.497.2018H} {497, 2018}

\bibitem[\protect\citeauthoryear{{Herpich}, {Stasi{\'n}ska}, {Mateus}, {Vale
  Asari}  \& {Cid Fernandes}}{{Herpich} et~al.}{2018}]{2018MNRAS.481.1774H}
{Herpich} F.,  {Stasi{\'n}ska} G.,  {Mateus} A.,  {Vale Asari} N.,   {Cid
  Fernandes} R.,  2018, \mn@doi [\mnras] {10.1093/mnras/sty2391}, \href
  {https://ui.adsabs.harvard.edu/abs/2018MNRAS.481.1774H} {481, 1774}

\bibitem[\protect\citeauthoryear{{Hood}, {Kannappan}, {Stark}, {Dell'Antonio},
  {Moffett}, {Eckert}, {Norris}  \& {Hendel}}{{Hood}
  et~al.}{2018}]{2018ApJ...857..144H}
{Hood} C.~E.,  {Kannappan} S.~J.,  {Stark} D.~V.,  {Dell'Antonio} I.~P.,
  {Moffett} A.~J.,  {Eckert} K.~D.,  {Norris} M.~A.,   {Hendel} D.,  2018,
  \mn@doi [\apj] {10.3847/1538-4357/aab719}, \href
  {https://ui.adsabs.harvard.edu/abs/2018ApJ...857..144H} {857, 144}

\bibitem[\protect\citeauthoryear{{Huang} \& {Fan}}{{Huang} \&
  {Fan}}{2022}]{2022ApJS..262...39H}
{Huang} Q.,  {Fan} L.,  2022, \mn@doi [\apjs] {10.3847/1538-4365/ac85b1}, \href
  {https://ui.adsabs.harvard.edu/abs/2022ApJS..262...39H} {262, 39}

\bibitem[\protect\citeauthoryear{{Ivezi{\'c}} et~al.,}{{Ivezi{\'c}}
  et~al.}{2019}]{2019ApJ...873..111I}
{Ivezi{\'c}} {\v{Z}}.,  et~al., 2019, \mn@doi [\apj]
  {10.3847/1538-4357/ab042c}, \href
  {https://ui.adsabs.harvard.edu/abs/2019ApJ...873..111I} {873, 111}

\bibitem[\protect\citeauthoryear{{Jesseit}, {Cappellari}, {Naab}, {Emsellem}
  \& {Burkert}}{{Jesseit} et~al.}{2009}]{2009MNRAS.397.1202J}
{Jesseit} R.,  {Cappellari} M.,  {Naab} T.,  {Emsellem} E.,   {Burkert} A.,
  2009, \mn@doi [\mnras] {10.1111/j.1365-2966.2009.14984.x}, \href
  {https://ui.adsabs.harvard.edu/abs/2009MNRAS.397.1202J} {397, 1202}

\bibitem[\protect\citeauthoryear{{Ji}, {Peirani}  \& {Yi}}{{Ji}
  et~al.}{2014}]{2014A&A...566A..97J}
{Ji} I.,  {Peirani} S.,   {Yi} S.~K.,  2014, \mn@doi [\aap]
  {10.1051/0004-6361/201423530}, \href
  {https://ui.adsabs.harvard.edu/abs/2014A&A...566A..97J} {566, A97}

\bibitem[\protect\citeauthoryear{{Johnston}, {Bullock}, {Sharma}, {Font},
  {Robertson}  \& {Leitner}}{{Johnston} et~al.}{2008}]{2008ApJ...689..936J}
{Johnston} K.~V.,  {Bullock} J.~S.,  {Sharma} S.,  {Font} A.,  {Robertson}
  B.~E.,   {Leitner} S.~N.,  2008, \mn@doi [\apj] {10.1086/592228}, \href
  {https://ui.adsabs.harvard.edu/abs/2008ApJ...689..936J} {689, 936}

\bibitem[\protect\citeauthoryear{{Kado-Fong} et~al.,}{{Kado-Fong}
  et~al.}{2018}]{2018ApJ...866..103K}
{Kado-Fong} E.,  et~al., 2018, \mn@doi [\apj] {10.3847/1538-4357/aae0f0}, \href
  {https://ui.adsabs.harvard.edu/abs/2018ApJ...866..103K} {866, 103}

\bibitem[\protect\citeauthoryear{{Karademir}, {Remus}, {Burkert}, {Dolag},
  {Hoffmann}, {Moster}, {Steinwandel}  \& {Zhang}}{{Karademir}
  et~al.}{2019}]{2019MNRAS.487..318K}
{Karademir} G.~S.,  {Remus} R.-S.,  {Burkert} A.,  {Dolag} K.,  {Hoffmann}
  T.~L.,  {Moster} B.~P.,  {Steinwandel} U.~P.,   {Zhang} J.,  2019, \mn@doi
  [\mnras] {10.1093/mnras/stz1251}, \href
  {https://ui.adsabs.harvard.edu/abs/2019MNRAS.487..318K} {487, 318}

\bibitem[\protect\citeauthoryear{{Kelvin} et~al.,}{{Kelvin}
  et~al.}{2014}]{2014MNRAS.439.1245K}
{Kelvin} L.~S.,  et~al., 2014, \mn@doi [\mnras] {10.1093/mnras/stt2391}, \href
  {https://ui.adsabs.harvard.edu/abs/2014MNRAS.439.1245K} {439, 1245}

\bibitem[\protect\citeauthoryear{{Kluge} et~al.,}{{Kluge}
  et~al.}{2020}]{2020ApJS..247...43K}
{Kluge} M.,  et~al., 2020, \mn@doi [\apjs] {10.3847/1538-4365/ab733b}, \href
  {https://ui.adsabs.harvard.edu/abs/2020ApJS..247...43K} {247, 43}

\bibitem[\protect\citeauthoryear{{Knapen}, {Cisternas}  \&
  {Querejeta}}{{Knapen} et~al.}{2015}]{2015MNRAS.454.1742K}
{Knapen} J.~H.,  {Cisternas} M.,   {Querejeta} M.,  2015, \mn@doi [\mnras]
  {10.1093/mnras/stv2135}, \href
  {https://ui.adsabs.harvard.edu/abs/2015MNRAS.454.1742K} {454, 1742}

\bibitem[\protect\citeauthoryear{{Kuijken} et~al.,}{{Kuijken}
  et~al.}{2019}]{2019A&A...625A...2K}
{Kuijken} K.,  et~al., 2019, \mn@doi [\aap] {10.1051/0004-6361/201834918},
  \href {https://ui.adsabs.harvard.edu/abs/2019A&A...625A...2K} {625, A2}

\bibitem[\protect\citeauthoryear{{Lagos}}{{Lagos}}{2020}]{2020IAUFM..30A.208L}
{Lagos} C. d.~P.,  2020, \mn@doi [IAU Focus Meeting]
  {10.1017/S1743921319004095}, \href
  {https://ui.adsabs.harvard.edu/abs/2020IAUFM..30A.208L} {30, 208}

\bibitem[\protect\citeauthoryear{{Lagos}, {Theuns}, {Stevens}, {Cortese},
  {Padilla}, {Davis}, {Contreras}  \& {Croton}}{{Lagos}
  et~al.}{2017}]{2017MNRAS.464.3850L}
{Lagos} C. d.~P.,  {Theuns} T.,  {Stevens} A. R.~H.,  {Cortese} L.,  {Padilla}
  N.~D.,  {Davis} T.~A.,  {Contreras} S.,   {Croton} D.,  2017, \mn@doi
  [\mnras] {10.1093/mnras/stw2610}, \href
  {https://ui.adsabs.harvard.edu/abs/2017MNRAS.464.3850L} {464, 3850}

\bibitem[\protect\citeauthoryear{{Lagos} et~al.,}{{Lagos}
  et~al.}{2018a}]{2018MNRAS.473.4956L}
{Lagos} C. d.~P.,  et~al., 2018a, \mn@doi [\mnras] {10.1093/mnras/stx2667},
  \href {https://ui.adsabs.harvard.edu/abs/2018MNRAS.473.4956L} {473, 4956}

\bibitem[\protect\citeauthoryear{{Lagos}, {Schaye}, {Bah{\'e}}, {van de Sande},
  {Kay}, {Barnes}, {Davis}  \& {Dalla Vecchia}}{{Lagos}
  et~al.}{2018b}]{2018MNRAS.476.4327L}
{Lagos} C. d.~P.,  {Schaye} J.,  {Bah{\'e}} Y.,  {van de Sande} J.,  {Kay}
  S.~T.,  {Barnes} D.,  {Davis} T.~A.,   {Dalla Vecchia} C.,  2018b, \mn@doi
  [\mnras] {10.1093/mnras/sty489}, \href
  {https://ui.adsabs.harvard.edu/abs/2018MNRAS.476.4327L} {476, 4327}

\bibitem[\protect\citeauthoryear{{Lagos}, {Emsellem}, {van de Sande},
  {Harborne}, {Cortese}, {Davison}, {Foster}  \& {Wright}}{{Lagos}
  et~al.}{2022}]{2022MNRAS.509.4372L}
{Lagos} C. d.~P.,  {Emsellem} E.,  {van de Sande} J.,  {Harborne} K.~E.,
  {Cortese} L.,  {Davison} T.,  {Foster} C.,   {Wright} R.~J.,  2022, \mn@doi
  [\mnras] {10.1093/mnras/stab3128}, \href
  {https://ui.adsabs.harvard.edu/abs/2022MNRAS.509.4372L} {509, 4372}

\bibitem[\protect\citeauthoryear{{Lee} et~al.,}{{Lee}
  et~al.}{2015}]{2015ApJ...801...80L}
{Lee} N.,  et~al., 2015, \mn@doi [\apj] {10.1088/0004-637X/801/2/80}, \href
  {https://ui.adsabs.harvard.edu/abs/2015ApJ...801...80L} {801, 80}

\bibitem[\protect\citeauthoryear{{Leslie} et~al.,}{{Leslie}
  et~al.}{2020}]{2020ApJ...899...58L}
{Leslie} S.~K.,  et~al., 2020, \mn@doi [\apj] {10.3847/1538-4357/aba044}, \href
  {https://ui.adsabs.harvard.edu/abs/2020ApJ...899...58L} {899, 58}

\bibitem[\protect\citeauthoryear{{Lofthouse}, {Kaviraj}, {Conselice},
  {Mortlock}  \& {Hartley}}{{Lofthouse} et~al.}{2017}]{2017MNRAS.465.2895L}
{Lofthouse} E.~K.,  {Kaviraj} S.,  {Conselice} C.~J.,  {Mortlock} A.,
  {Hartley} W.,  2017, \mn@doi [\mnras] {10.1093/mnras/stw2895}, \href
  {https://ui.adsabs.harvard.edu/abs/2017MNRAS.465.2895L} {465, 2895}

\bibitem[\protect\citeauthoryear{{L{\'o}pez-S{\'a}nchez}}{{L{\'o}pez-S{\'a}nchez}}{2010}]{2010A&A...521A..63L}
{L{\'o}pez-S{\'a}nchez} {\'A}.~R.,  2010, \mn@doi [\aap]
  {10.1051/0004-6361/201014295}, \href
  {https://ui.adsabs.harvard.edu/abs/2010A&A...521A..63L} {521, A63}

\bibitem[\protect\citeauthoryear{{Lotz}, {Jonsson}, {Cox}  \& {Primack}}{{Lotz}
  et~al.}{2008}]{2008MNRAS.391.1137L}
{Lotz} J.~M.,  {Jonsson} P.,  {Cox} T.~J.,   {Primack} J.~R.,  2008, \mn@doi
  [\mnras] {10.1111/j.1365-2966.2008.14004.x}, \href
  {https://ui.adsabs.harvard.edu/abs/2008MNRAS.391.1137L} {391, 1137}

\bibitem[\protect\citeauthoryear{{Lotz}, {Jonsson}, {Cox}  \& {Primack}}{{Lotz}
  et~al.}{2010a}]{2010MNRAS.404..575L}
{Lotz} J.~M.,  {Jonsson} P.,  {Cox} T.~J.,   {Primack} J.~R.,  2010a, \mn@doi
  [\mnras] {10.1111/j.1365-2966.2010.16268.x}, \href
  {https://ui.adsabs.harvard.edu/abs/2010MNRAS.404..575L} {404, 575}

\bibitem[\protect\citeauthoryear{{Lotz}, {Jonsson}, {Cox}  \& {Primack}}{{Lotz}
  et~al.}{2010b}]{2010MNRAS.404..590L}
{Lotz} J.~M.,  {Jonsson} P.,  {Cox} T.~J.,   {Primack} J.~R.,  2010b, \mn@doi
  [\mnras] {10.1111/j.1365-2966.2010.16269.x}, \href
  {https://ui.adsabs.harvard.edu/abs/2010MNRAS.404..590L} {404, 590}

\bibitem[\protect\citeauthoryear{{Mancillas}, {Duc}, {Combes}, {Bournaud},
  {Emsellem}, {Martig}  \& {Michel-Dansac}}{{Mancillas}
  et~al.}{2019}]{2019A&A...632A.122M}
{Mancillas} B.,  {Duc} P.-A.,  {Combes} F.,  {Bournaud} F.,  {Emsellem} E.,
  {Martig} M.,   {Michel-Dansac} L.,  2019, \mn@doi [\aap]
  {10.1051/0004-6361/201936320}, \href
  {https://ui.adsabs.harvard.edu/abs/2019A&A...632A.122M} {632, A122}

\bibitem[\protect\citeauthoryear{{Mantha} et~al.,}{{Mantha}
  et~al.}{2019}]{2019MNRAS.486.2643M}
{Mantha} K.~B.,  et~al., 2019, \mn@doi [\mnras] {10.1093/mnras/stz872}, \href
  {https://ui.adsabs.harvard.edu/abs/2019MNRAS.486.2643M} {486, 2643}

\bibitem[\protect\citeauthoryear{{Martin} et~al.,}{{Martin}
  et~al.}{2022}]{2022MNRAS.513.1459M}
{Martin} G.,  et~al., 2022, \mn@doi [\mnras] {10.1093/mnras/stac1003}, \href
  {https://ui.adsabs.harvard.edu/abs/2022MNRAS.513.1459M} {513, 1459}

\bibitem[\protect\citeauthoryear{{McDermid} et~al.,}{{McDermid}
  et~al.}{2015}]{2015MNRAS.448.3484M}
{McDermid} R.~M.,  et~al., 2015, \mn@doi [\mnras] {10.1093/mnras/stv105}, \href
  {https://ui.adsabs.harvard.edu/abs/2015MNRAS.448.3484M} {448, 3484}

\bibitem[\protect\citeauthoryear{{McIntosh}, {Guo}, {Hertzberg}, {Katz}, {Mo},
  {van den Bosch}  \& {Yang}}{{McIntosh} et~al.}{2008}]{2008MNRAS.388.1537M}
{McIntosh} D.~H.,  {Guo} Y.,  {Hertzberg} J.,  {Katz} N.,  {Mo} H.~J.,  {van
  den Bosch} F.~C.,   {Yang} X.,  2008, \mn@doi [\mnras]
  {10.1111/j.1365-2966.2008.13531.x}, \href
  {https://ui.adsabs.harvard.edu/abs/2008MNRAS.388.1537M} {388, 1537}

\bibitem[\protect\citeauthoryear{{Miyazaki} et~al.,}{{Miyazaki}
  et~al.}{2015}]{2015ApJ...807...22M}
{Miyazaki} S.,  et~al., 2015, \mn@doi [\apj] {10.1088/0004-637X/807/1/22},
  \href {https://ui.adsabs.harvard.edu/abs/2015ApJ...807...22M} {807, 22}

\bibitem[\protect\citeauthoryear{{Naab} et~al.,}{{Naab}
  et~al.}{2014}]{2014MNRAS.444.3357N}
{Naab} T.,  et~al., 2014, \mn@doi [\mnras] {10.1093/mnras/stt1919}, \href
  {https://ui.adsabs.harvard.edu/abs/2014MNRAS.444.3357N} {444, 3357}

\bibitem[\protect\citeauthoryear{{Nevin}, {Blecha}, {Comerford}  \&
  {Greene}}{{Nevin} et~al.}{2019}]{2019ApJ...872...76N}
{Nevin} R.,  {Blecha} L.,  {Comerford} J.,   {Greene} J.,  2019, \mn@doi [\apj]
  {10.3847/1538-4357/aafd34}, \href
  {https://ui.adsabs.harvard.edu/abs/2019ApJ...872...76N} {872, 76}

\bibitem[\protect\citeauthoryear{{Nevin} et~al.,}{{Nevin}
  et~al.}{2021}]{2021ApJ...912...45N}
{Nevin} R.,  et~al., 2021, \mn@doi [\apj] {10.3847/1538-4357/abe2a9}, \href
  {https://ui.adsabs.harvard.edu/abs/2021ApJ...912...45N} {912, 45}

\bibitem[\protect\citeauthoryear{{Oh}, {Kim}, {Lee}  \& {Kim}}{{Oh}
  et~al.}{2008}]{2008ApJ...683...94O}
{Oh} S.~H.,  {Kim} W.-T.,  {Lee} H.~M.,   {Kim} J.,  2008, \mn@doi [\apj]
  {10.1086/588184}, \href
  {https://ui.adsabs.harvard.edu/abs/2008ApJ...683...94O} {683, 94}

\bibitem[\protect\citeauthoryear{{Oh} et~al.,}{{Oh}
  et~al.}{2016}]{2016ApJ...832...69O}
{Oh} S.,  et~al., 2016, \mn@doi [\apj] {10.3847/0004-637X/832/1/69}, \href
  {https://ui.adsabs.harvard.edu/abs/2016ApJ...832...69O} {832, 69}

\bibitem[\protect\citeauthoryear{{Oke} \& {Gunn}}{{Oke} \&
  {Gunn}}{1983}]{1983ApJ...266..713O}
{Oke} J.~B.,  {Gunn} J.~E.,  1983, \mn@doi [\apj] {10.1086/160817}, \href
  {https://ui.adsabs.harvard.edu/abs/1983ApJ...266..713O} {266, 713}

\bibitem[\protect\citeauthoryear{{Owers} et~al.,}{{Owers}
  et~al.}{2017}]{2017MNRAS.468.1824O}
{Owers} M.~S.,  et~al., 2017, \mn@doi [\mnras] {10.1093/mnras/stx562}, \href
  {https://ui.adsabs.harvard.edu/abs/2017MNRAS.468.1824O} {468, 1824}

\bibitem[\protect\citeauthoryear{{Peng}, {Ho}, {Impey}  \& {Rix}}{{Peng}
  et~al.}{2002}]{2002AJ....124..266P}
{Peng} C.~Y.,  {Ho} L.~C.,  {Impey} C.~D.,   {Rix} H.-W.,  2002, \mn@doi [\aj]
  {10.1086/340952}, \href
  {https://ui.adsabs.harvard.edu/abs/2002AJ....124..266P} {124, 266}

\bibitem[\protect\citeauthoryear{{Penoyre}, {Moster}, {Sijacki}  \&
  {Genel}}{{Penoyre} et~al.}{2017}]{2017MNRAS.468.3883P}
{Penoyre} Z.,  {Moster} B.~P.,  {Sijacki} D.,   {Genel} S.,  2017, \mn@doi
  [\mnras] {10.1093/mnras/stx762}, \href
  {https://ui.adsabs.harvard.edu/abs/2017MNRAS.468.3883P} {468, 3883}

\bibitem[\protect\citeauthoryear{{Pietrinferni}, {Cassisi}, {Salaris}  \&
  {Castelli}}{{Pietrinferni} et~al.}{2004}]{2004ApJ...612..168P}
{Pietrinferni} A.,  {Cassisi} S.,  {Salaris} M.,   {Castelli} F.,  2004,
  \mn@doi [\apj] {10.1086/422498}, \href
  {https://ui.adsabs.harvard.edu/abs/2004ApJ...612..168P} {612, 168}

\bibitem[\protect\citeauthoryear{{Pietrinferni}, {Cassisi}, {Salaris}  \&
  {Castelli}}{{Pietrinferni} et~al.}{2006}]{2006ApJ...642..797P}
{Pietrinferni} A.,  {Cassisi} S.,  {Salaris} M.,   {Castelli} F.,  2006,
  \mn@doi [\apj] {10.1086/501344}, \href
  {https://ui.adsabs.harvard.edu/abs/2006ApJ...642..797P} {642, 797}

\bibitem[\protect\citeauthoryear{{Planck Collaboration} et~al.,}{{Planck
  Collaboration} et~al.}{2016}]{2016A&A...594A..13P}
{Planck Collaboration} et~al., 2016, \mn@doi [\aap]
  {10.1051/0004-6361/201525830}, \href
  {https://ui.adsabs.harvard.edu/abs/2016A&A...594A..13P} {594, A13}

\bibitem[\protect\citeauthoryear{{Pontzen}, {Tremmel}, {Roth}, {Peiris},
  {Saintonge}, {Volonteri}, {Quinn}  \& {Governato}}{{Pontzen}
  et~al.}{2017}]{2017MNRAS.465..547P}
{Pontzen} A.,  {Tremmel} M.,  {Roth} N.,  {Peiris} H.~V.,  {Saintonge} A.,
  {Volonteri} M.,  {Quinn} T.,   {Governato} F.,  2017, \mn@doi [\mnras]
  {10.1093/mnras/stw2627}, \href
  {https://ui.adsabs.harvard.edu/abs/2017MNRAS.465..547P} {465, 547}

\bibitem[\protect\citeauthoryear{{Pop}, {Pillepich}, {Amorisco}  \&
  {Hernquist}}{{Pop} et~al.}{2018}]{2018MNRAS.480.1715P}
{Pop} A.-R.,  {Pillepich} A.,  {Amorisco} N.~C.,   {Hernquist} L.,  2018,
  \mn@doi [\mnras] {10.1093/mnras/sty1932}, \href
  {https://ui.adsabs.harvard.edu/abs/2018MNRAS.480.1715P} {480, 1715}

\bibitem[\protect\citeauthoryear{{Ristea} et~al.,}{{Ristea}
  et~al.}{2022}]{2022MNRAS.517.2677R}
{Ristea} A.,  et~al., 2022, \mn@doi [\mnras] {10.1093/mnras/stac2839}, \href
  {https://ui.adsabs.harvard.edu/abs/2022MNRAS.517.2677R} {517, 2677}

\bibitem[\protect\citeauthoryear{{Robotham}, {Taranu}, {Tobar}, {Moffett}  \&
  {Driver}}{{Robotham} et~al.}{2017}]{2017MNRAS.466.1513R}
{Robotham} A.~S.~G.,  {Taranu} D.~S.,  {Tobar} R.,  {Moffett} A.,   {Driver}
  S.~P.,  2017, \mn@doi [\mnras] {10.1093/mnras/stw3039}, \href
  {https://ui.adsabs.harvard.edu/abs/2017MNRAS.466.1513R} {466, 1513}

\bibitem[\protect\citeauthoryear{{Robotham}, {Davies}, {Driver}, {Koushan},
  {Taranu}, {Casura}  \& {Liske}}{{Robotham}
  et~al.}{2018}]{2018MNRAS.476.3137R}
{Robotham} A.~S.~G.,  {Davies} L.~J.~M.,  {Driver} S.~P.,  {Koushan} S.,
  {Taranu} D.~S.,  {Casura} S.,   {Liske} J.,  2018, \mn@doi [\mnras]
  {10.1093/mnras/sty440}, \href
  {https://ui.adsabs.harvard.edu/abs/2018MNRAS.476.3137R} {476, 3137}

\bibitem[\protect\citeauthoryear{{Sales}, {Navarro}, {Theuns}, {Schaye},
  {White}, {Frenk}, {Crain}  \& {Dalla Vecchia}}{{Sales}
  et~al.}{2012}]{2012MNRAS.423.1544S}
{Sales} L.~V.,  {Navarro} J.~F.,  {Theuns} T.,  {Schaye} J.,  {White} S. D.~M.,
   {Frenk} C.~S.,  {Crain} R.~A.,   {Dalla Vecchia} C.,  2012, \mn@doi [\mnras]
  {10.1111/j.1365-2966.2012.20975.x}, \href
  {https://ui.adsabs.harvard.edu/abs/2012MNRAS.423.1544S} {423, 1544}

\bibitem[\protect\citeauthoryear{{Schulze}, {Remus}, {Dolag}, {Bellstedt},
  {Burkert}  \& {Forbes}}{{Schulze} et~al.}{2020}]{2020MNRAS.493.3778S}
{Schulze} F.,  {Remus} R.-S.,  {Dolag} K.,  {Bellstedt} S.,  {Burkert} A.,
  {Forbes} D.~A.,  2020, \mn@doi [\mnras] {10.1093/mnras/staa511}, \href
  {https://ui.adsabs.harvard.edu/abs/2020MNRAS.493.3778S} {493, 3778}

\bibitem[\protect\citeauthoryear{{Schweizer} \& {Seitzer}}{{Schweizer} \&
  {Seitzer}}{1988}]{1988ApJ...328...88S}
{Schweizer} F.,  {Seitzer} P.,  1988, \mn@doi [\apj] {10.1086/166270}, \href
  {https://ui.adsabs.harvard.edu/abs/1988ApJ...328...88S} {328, 88}

\bibitem[\protect\citeauthoryear{{Scott} \& {Kaviraj}}{{Scott} \&
  {Kaviraj}}{2014}]{2014MNRAS.437.2137S}
{Scott} C.,  {Kaviraj} S.,  2014, \mn@doi [\mnras] {10.1093/mnras/stt2014},
  \href {https://ui.adsabs.harvard.edu/abs/2014MNRAS.437.2137S} {437, 2137}

\bibitem[\protect\citeauthoryear{{Scott} et~al.,}{{Scott}
  et~al.}{2009}]{2009MNRAS.398.1835S}
{Scott} N.,  et~al., 2009, \mn@doi [\mnras] {10.1111/j.1365-2966.2009.15275.x},
  \href {https://ui.adsabs.harvard.edu/abs/2009MNRAS.398.1835S} {398, 1835}

\bibitem[\protect\citeauthoryear{{Scott} et~al.,}{{Scott}
  et~al.}{2018}]{2018MNRAS.481.2299S}
{Scott} N.,  et~al., 2018, \mn@doi [\mnras] {10.1093/mnras/sty2355}, \href
  {https://ui.adsabs.harvard.edu/abs/2018MNRAS.481.2299S} {481, 2299}

\bibitem[\protect\citeauthoryear{{Sharp} et~al.,}{{Sharp}
  et~al.}{2006}]{2006SPIE.6269E..0GS}
{Sharp} R.,  et~al., 2006, in {McLean} I.~S.,  {Iye} M.,  eds,  Society of
  Photo-Optical Instrumentation Engineers (SPIE) Conference Series Vol. 6269,
  Ground-based and Airborne Instrumentation for Astronomy. p. 62690G
  (\mn@eprint {arXiv} {astro-ph/0606137}), \mn@doi{10.1117/12.671022}

\bibitem[\protect\citeauthoryear{{Sharp} et~al.,}{{Sharp}
  et~al.}{2015}]{2015MNRAS.446.1551S}
{Sharp} R.,  et~al., 2015, \mn@doi [\mnras] {10.1093/mnras/stu2055}, \href
  {https://ui.adsabs.harvard.edu/abs/2015MNRAS.446.1551S} {446, 1551}

\bibitem[\protect\citeauthoryear{{Sola} et~al.,}{{Sola}
  et~al.}{2022}]{2022A&A...662A.124S}
{Sola} E.,  et~al., 2022, \mn@doi [\aap] {10.1051/0004-6361/202142675}, \href
  {https://ui.adsabs.harvard.edu/abs/2022A&A...662A.124S} {662, A124}

\bibitem[\protect\citeauthoryear{{Sparre} \& {Springel}}{{Sparre} \&
  {Springel}}{2017}]{2017MNRAS.470.3946S}
{Sparre} M.,  {Springel} V.,  2017, \mn@doi [\mnras] {10.1093/mnras/stx1516},
  \href {https://ui.adsabs.harvard.edu/abs/2017MNRAS.470.3946S} {470, 3946}

\bibitem[\protect\citeauthoryear{{Tal}, {van Dokkum}, {Nelan}  \&
  {Bezanson}}{{Tal} et~al.}{2009}]{2009AJ....138.1417T}
{Tal} T.,  {van Dokkum} P.~G.,  {Nelan} J.,   {Bezanson} R.,  2009, \mn@doi
  [\aj] {10.1088/0004-6256/138/5/1417}, \href
  {https://ui.adsabs.harvard.edu/abs/2009AJ....138.1417T} {138, 1417}

\bibitem[\protect\citeauthoryear{{Taylor} et~al.,}{{Taylor}
  et~al.}{2011}]{2011MNRAS.418.1587T}
{Taylor} E.~N.,  et~al., 2011, \mn@doi [\mnras]
  {10.1111/j.1365-2966.2011.19536.x}, \href
  {https://ui.adsabs.harvard.edu/abs/2011MNRAS.418.1587T} {418, 1587}

\bibitem[\protect\citeauthoryear{{Teklu}, {Remus}, {Dolag}, {Beck}, {Burkert},
  {Schmidt}, {Schulze}  \& {Steinborn}}{{Teklu}
  et~al.}{2015}]{2015ApJ...812...29T}
{Teklu} A.~F.,  {Remus} R.-S.,  {Dolag} K.,  {Beck} A.~M.,  {Burkert} A.,
  {Schmidt} A.~S.,  {Schulze} F.,   {Steinborn} L.~K.,  2015, \mn@doi [\apj]
  {10.1088/0004-637X/812/1/29}, \href
  {https://ui.adsabs.harvard.edu/abs/2015ApJ...812...29T} {812, 29}

\bibitem[\protect\citeauthoryear{{Thorp}, {Ellison}, {Simard}, {S{\'a}nchez}
  \& {Antonio}}{{Thorp} et~al.}{2019}]{2019MNRAS.482L..55T}
{Thorp} M.~D.,  {Ellison} S.~L.,  {Simard} L.,  {S{\'a}nchez} S.~F.,
  {Antonio} B.,  2019, \mn@doi [\mnras] {10.1093/mnrasl/sly185}, \href
  {https://ui.adsabs.harvard.edu/abs/2019MNRAS.482L..55T} {482, L55}

\bibitem[\protect\citeauthoryear{{Toomre} \& {Toomre}}{{Toomre} \&
  {Toomre}}{1972}]{1972ApJ...178..623T}
{Toomre} A.,  {Toomre} J.,  1972, \mn@doi [\apj] {10.1086/151823}, \href
  {https://ui.adsabs.harvard.edu/abs/1972ApJ...178..623T} {178, 623}

\bibitem[\protect\citeauthoryear{{Valenzuela} \& {Remus}}{{Valenzuela} \&
  {Remus}}{2022}]{2022arXiv220808443V}
{Valenzuela} L.~M.,  {Remus} R.-S.,  2022, \mn@doi [arXiv e-prints]
  {10.48550/arXiv.2208.08443}, \href
  {https://ui.adsabs.harvard.edu/abs/2022arXiv220808443V} {p. arXiv:2208.08443}

\bibitem[\protect\citeauthoryear{{Vaughan} et~al.,}{{Vaughan}
  et~al.}{2022}]{2022MNRAS.516.2971V}
{Vaughan} S.~P.,  et~al., 2022, \mn@doi [\mnras] {10.1093/mnras/stac2304},
  \href {https://ui.adsabs.harvard.edu/abs/2022MNRAS.516.2971V} {516, 2971}

\bibitem[\protect\citeauthoryear{{Vazdekis} et~al.,}{{Vazdekis}
  et~al.}{2015}]{2015MNRAS.449.1177V}
{Vazdekis} A.,  et~al., 2015, \mn@doi [\mnras] {10.1093/mnras/stv151}, \href
  {https://ui.adsabs.harvard.edu/abs/2015MNRAS.449.1177V} {449, 1177}

\bibitem[\protect\citeauthoryear{{Virtanen} et~al.,}{{Virtanen}
  et~al.}{2020}]{2020NatMe..17..261V}
{Virtanen} P.,  et~al., 2020, \mn@doi [Nature Methods]
  {10.1038/s41592-019-0686-2}, \href
  {https://ui.adsabs.harvard.edu/abs/2020NatMe..17..261V} {17, 261}

\bibitem[\protect\citeauthoryear{{Walo-Mart{\'\i}n}, {Falc{\'o}n-Barroso},
  {Dalla Vecchia}, {P{\'e}rez}  \& {Negri}}{{Walo-Mart{\'\i}n}
  et~al.}{2020}]{2020MNRAS.494.5652W}
{Walo-Mart{\'\i}n} D.,  {Falc{\'o}n-Barroso} J.,  {Dalla Vecchia} C.,
  {P{\'e}rez} I.,   {Negri} A.,  2020, \mn@doi [\mnras]
  {10.1093/mnras/staa1066}, \href
  {https://ui.adsabs.harvard.edu/abs/2020MNRAS.494.5652W} {494, 5652}

\bibitem[\protect\citeauthoryear{{Weil}, {Bland-Hawthorn}  \& {Malin}}{{Weil}
  et~al.}{1997}]{1997ApJ...490..664W}
{Weil} M.~L.,  {Bland-Hawthorn} J.,   {Malin} D.~F.,  1997, \mn@doi [\apj]
  {10.1086/304886}, \href
  {https://ui.adsabs.harvard.edu/abs/1997ApJ...490..664W} {490, 664}

\bibitem[\protect\citeauthoryear{{White} \& {Rees}}{{White} \&
  {Rees}}{1978}]{1978MNRAS.183..341W}
{White} S.~D.~M.,  {Rees} M.~J.,  1978, \mn@doi [\mnras]
  {10.1093/mnras/183.3.341}, \href
  {https://ui.adsabs.harvard.edu/abs/1978MNRAS.183..341W} {183, 341}

\bibitem[\protect\citeauthoryear{{Wilman}, {Fontanot}, {De Lucia}, {Erwin}  \&
  {Monaco}}{{Wilman} et~al.}{2013}]{2013MNRAS.433.2986W}
{Wilman} D.~J.,  {Fontanot} F.,  {De Lucia} G.,  {Erwin} P.,   {Monaco} P.,
  2013, \mn@doi [\mnras] {10.1093/mnras/stt941}, \href
  {https://ui.adsabs.harvard.edu/abs/2013MNRAS.433.2986W} {433, 2986}

\bibitem[\protect\citeauthoryear{{van Dokkum}}{{van
  Dokkum}}{2005}]{2005AJ....130.2647V}
{van Dokkum} P.~G.,  2005, \mn@doi [\aj] {10.1086/497593}, \href
  {https://ui.adsabs.harvard.edu/abs/2005AJ....130.2647V} {130, 2647}

\bibitem[\protect\citeauthoryear{{van de Sande} et~al.,}{{van de Sande}
  et~al.}{2017a}]{2017MNRAS.472.1272V}
{van de Sande} J.,  et~al., 2017a, \mn@doi [\mnras] {10.1093/mnras/stx1751},
  \href {https://ui.adsabs.harvard.edu/abs/2017MNRAS.472.1272V} {472, 1272}

\bibitem[\protect\citeauthoryear{{van de Sande} et~al.,}{{van de Sande}
  et~al.}{2017b}]{2017ApJ...835..104V}
{van de Sande} J.,  et~al., 2017b, \mn@doi [\apj]
  {10.3847/1538-4357/835/1/104}, \href
  {https://ui.adsabs.harvard.edu/abs/2017ApJ...835..104V} {835, 104}

\bibitem[\protect\citeauthoryear{{van de Sande} et~al.,}{{van de Sande}
  et~al.}{2018}]{2018NatAs...2..483V}
{van de Sande} J.,  et~al., 2018, \mn@doi [Nature Astronomy]
  {10.1038/s41550-018-0436-x}, \href
  {https://ui.adsabs.harvard.edu/abs/2018NatAs...2..483V} {2, 483}

\bibitem[\protect\citeauthoryear{{van de Sande} et~al.,}{{van de Sande}
  et~al.}{2021}]{2021MNRAS.505.3078V}
{van de Sande} J.,  et~al., 2021, \mn@doi [\mnras] {10.1093/mnras/stab1490},
  \href {https://ui.adsabs.harvard.edu/abs/2021MNRAS.505.3078V} {505, 3078}

\makeatother
\end{thebibliography}




\appendix

\section{Stellar Mass, Age, \lre and Tidal Features}

\update{In this section we show that although stellar mass is strongly correlated with tidal features, it does not drive the relationship between \lre, mean stellar age and shells seen in Figure \ref{fig:spin_age}.}

\begin{figure*}
	\includegraphics[width=0.95\textwidth]{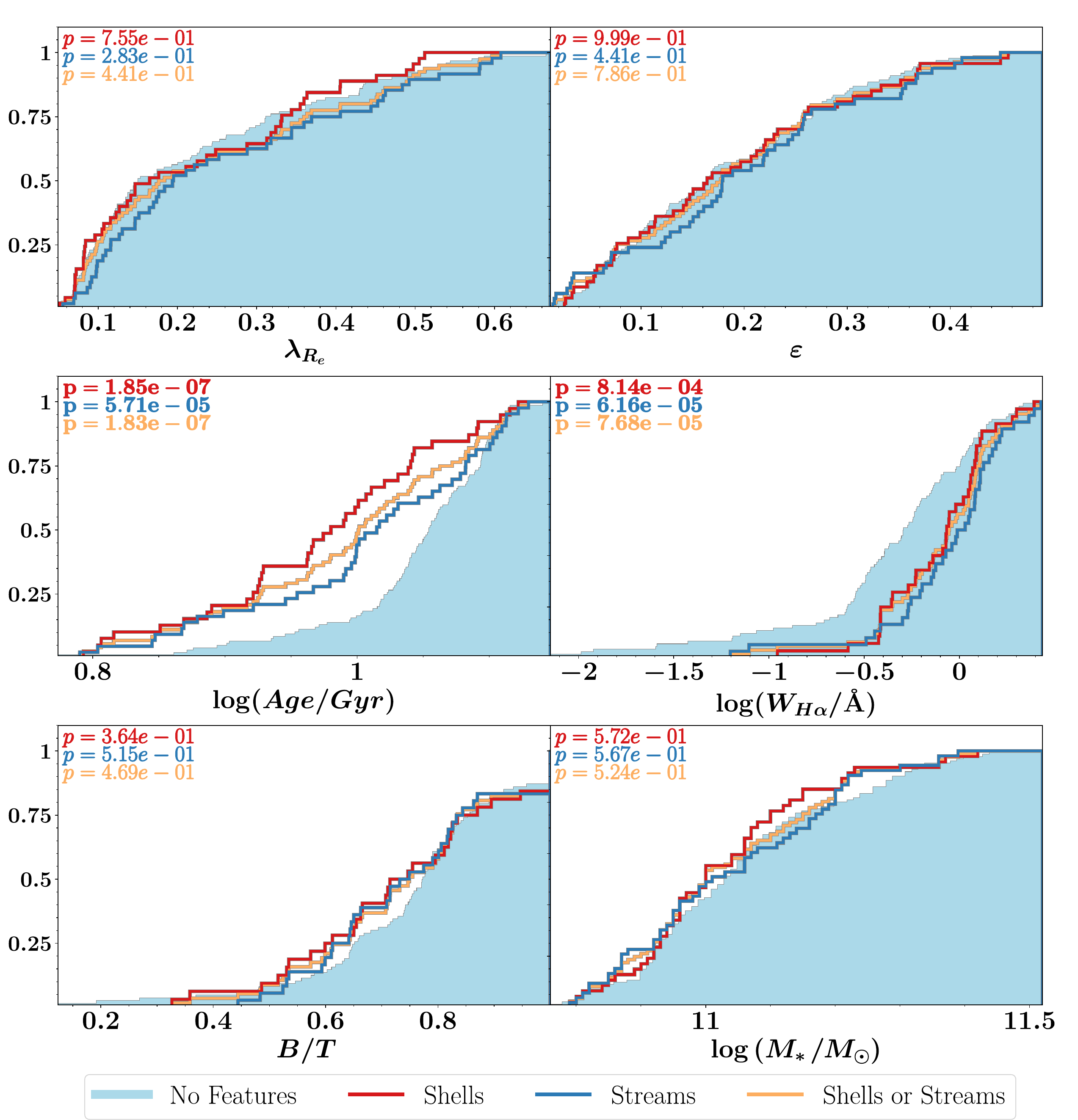}
    \caption{The cumulative distribution for all relevant parameters and tidal feature samples, for galaxies with stellar masses above the median stellar mass ($\logm = 10.75$). All cumulative distributions and labels are the same as Figure \ref{fig:dists}. We find that for these galaxies with high stellar mass, there is no correlation between stellar mass and tidal features.}
    \label{fig:cum_dists_all_highmass}
\end{figure*}

\update{In Figure \ref{fig:cum_dists_all_highmass}, we see that when we restrict to galaxies with a stellar mass above the median ($\logm = 10.75$), stellar mass is no longer correlated with tidal features. As discussed in Section \ref{sec:radial_infall_mergers}, we still see a significant split in feature fraction with age at this mass. Additionally, the same qualitative results from Figure \ref{fig:spin_age} are found for this mass group.}

\update{We further show in Figure \ref{fig:cum_dists_all_highmass} that the correlation between shells and age becomes \textit{stronger} when we restrict to higher stellar masses, as also seen in panel (b) of Figure \ref{fig:all_parms}. Although there are a high number of shells at all ages at high mass, the fraction of shells is clearly much higher at low stellar ages.}

\section{Example Shells}

\begin{figure*}
	\includegraphics[width=0.95\textwidth]{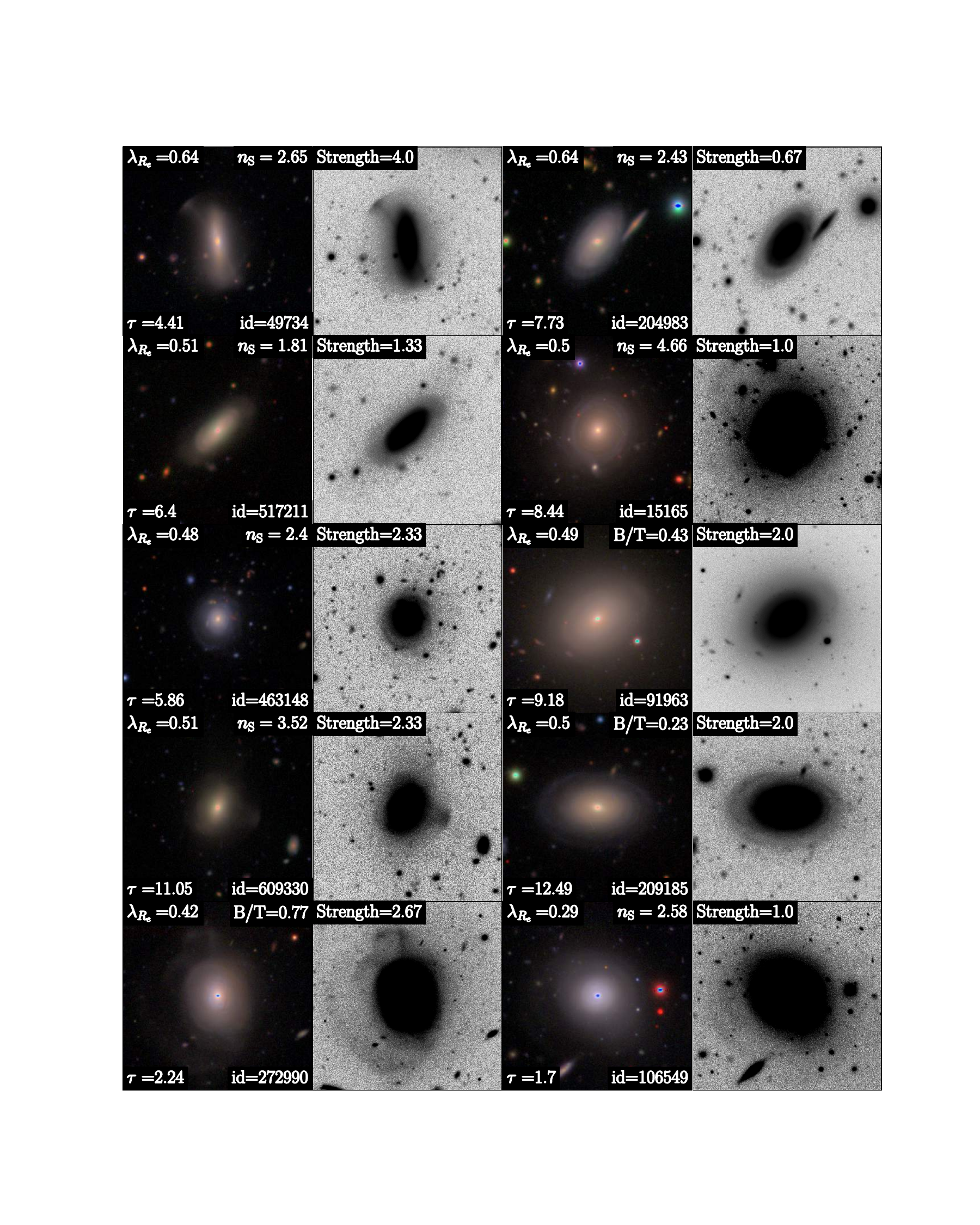}
    \caption{Galaxies which were both classified as containing a shell, and have relatively high \lre\ or low mean light-weighted stellar age. Each black and white cutout corresponds to the colour image to its left. We include values for the galaxy's \lre, age ($\tau$, measured in Gyr), CATAID (id), B/T (or Sérsic index if B/T is not available), and its classified feature strength.}
    \label{fig:interesting_gals}
\end{figure*}

Here we present examples of shell galaxy cutouts with colour images as well. In Figure \ref{fig:interesting_gals}, we show galaxies which despite being classified as having a shell, display relatively high \lre\ or low light-weighted mean stellar age. These are examples of galaxies in which it is possible that a ring or weak spiral arm was mis-characterised as a shell.

\begin{figure*}
	\includegraphics[width=\textwidth]{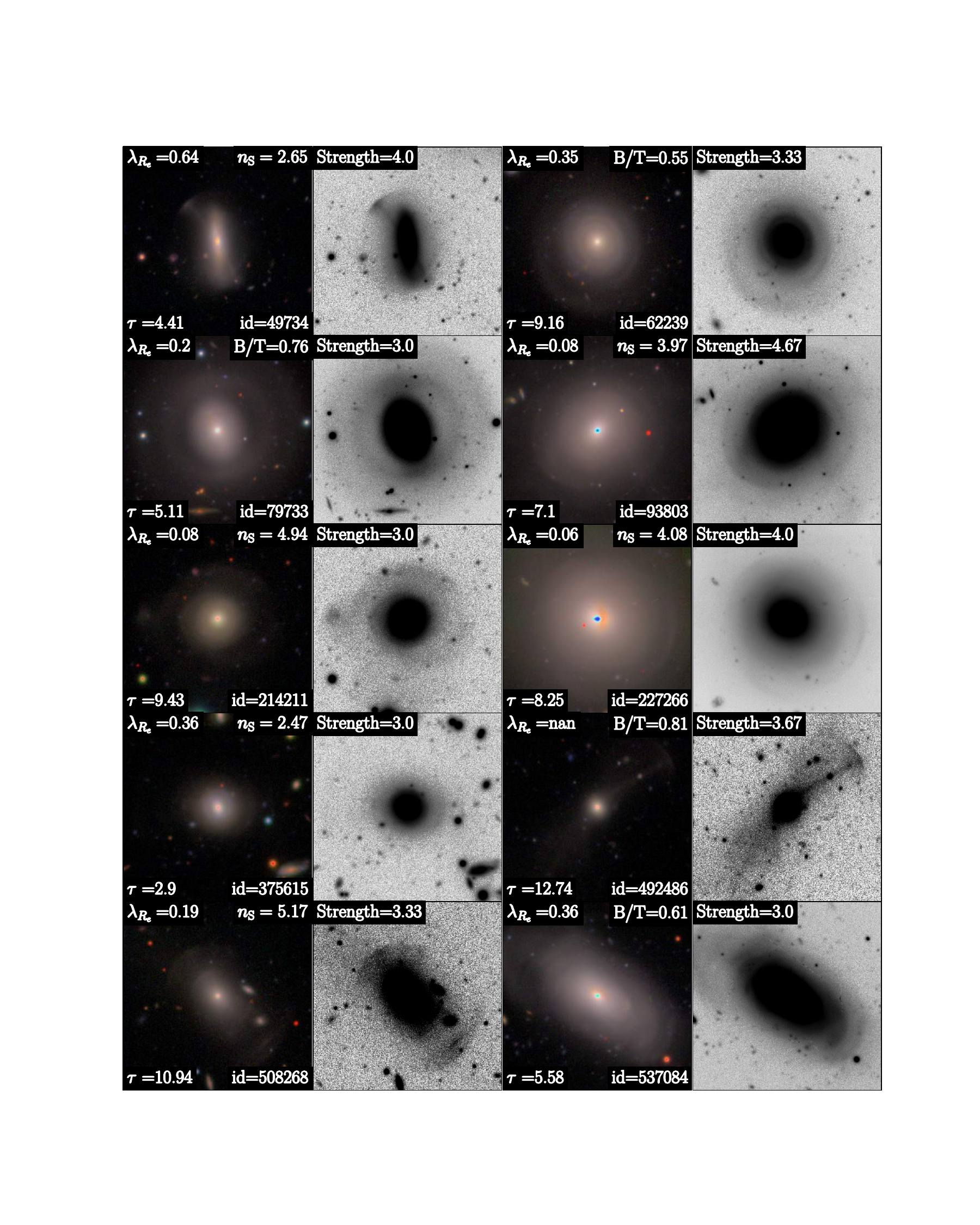}
    \caption{Galaxies which were both classified as containing a shell, and have a classified strength of at least 3/5. The image layout is similar to Figure \ref{fig:interesting_gals}.}
    \label{fig:strong_shells}
\end{figure*}

In Figure \ref{fig:strong_shells}, we show examples galaxies which were classified as having a strong shell, i.e. a shell strength of at least 3/5.

\section{Magneticum ETG Definition}

\begin{figure*}
	\includegraphics[width=\textwidth]{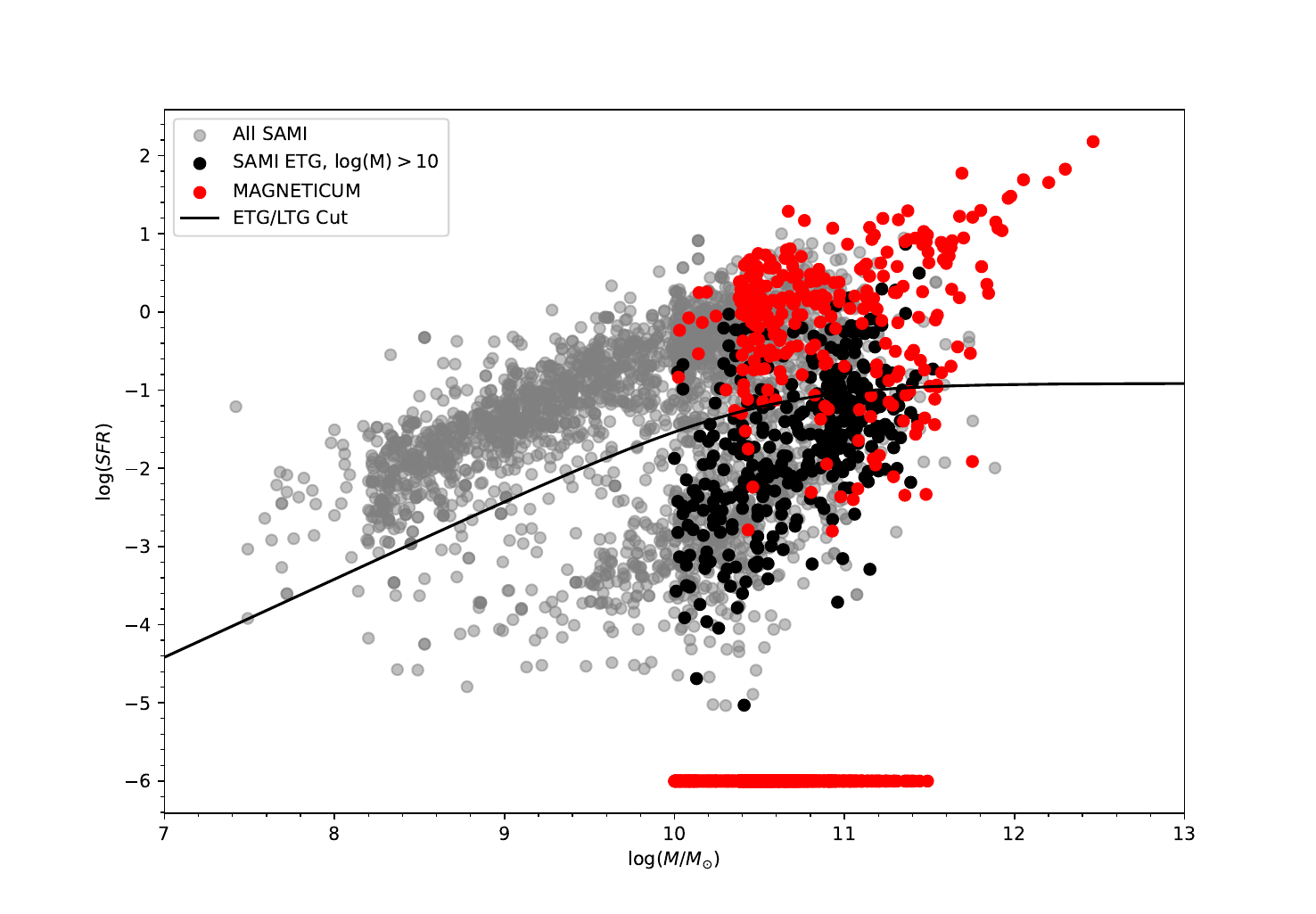}
    \caption{The star forming main sequence for SAMI galaxies, with Magneticum overplotted. We make a cut based on the SAMI ETGs in order to defne Magneticum ETGs, with all Magneticum galaxies below the line being classified as ETGs. The cut is based on a parametrised form taken from \citet{2020ApJ...899...58L}. The majority of Magneticum ETGs can be seen to have $\log(SFR)=-6$, essentially entirely passive galaxies.}
    \label{fig:magneticum_etg_cut}
\end{figure*}

Here we show the cut in the star forming main sequence of Magneticum galaxies, used to classify them as either ETGs or LTGs. In Figure \ref{fig:magneticum_etg_cut}, we show all SAMI galaxies, and our ETG sample on the SAMI star forming main sequence. We then make a cut on Magneticum galaxies, based on a fit to our star forming main sequence, shifted down in SFR by 1 dex. The functional form (shown in Equations \ref{eq:sfr_func_form_1} and \ref{eq:sfr_func_form_2}) was used by \citet{2021MNRAS.503.4992F} and \citet{2020ApJ...899...58L}, inspired by \citet{2015ApJ...801...80L}.
\begin{align}
    \label{eq:sfr_func_form_1}
    \log(SFR)&=S_0-a_1t-\log\bigg(1+\bigg(\frac{10^{\textit{M}'_t}}{10^{\textit{M}}}\bigg)\bigg),\\
    \label{eq:sfr_func_form_2}
    \textit{M}'_t&=M_0-a_2t
\end{align}
where $t$ is the age of the universe (taken to be 13.5 Gyr for SAMI data by \citet{2021MNRAS.503.4992F}) and \textit{M} is \logm. $S_0$, $a_1$, $a_2$ and $M_0$ are parameters which are fit\footnote{Parameters are fit using a linear least squares method \citep{2020NatMe..17..261V}}. Our best fit can be seen in Equations \ref{eq:sfr_fit_1} and \ref{eq:sfr_fit_2}.
\begin{align}
    \label{eq:sfr_fit_1}
    \log(SFR)&=2.91-0.209\times13.5-\log\bigg(1+\bigg(\frac{10^{\textit{M}'_t}}{10^{\textit{M}}}\bigg)\bigg),\\
    \label{eq:sfr_fit_2}
    \textit{M}'_t&=13.201-0.200\times13.5
\end{align}
This was then shifted down by 1 dex, and can be seen in Figure \ref{fig:magneticum_etg_cut}.

\bsp	
\label{lastpage}
\end{document}